\RequirePackage{fix-cm}
\documentclass[natbib,smallextended]{svjour3} % onecolumn (second format)

\smartqed  % flush right qed marks, e.g. at end of proof

\usepackage{graphicx}%
\usepackage{multirow}
\usepackage{aps-bibstyle}  % use this style if you upload to .tex file only a part of Bibtex created bbl.

\usepackage{amssymb}
\usepackage{amsmath}
\usepackage{color}
\usepackage{xspace}
\usepackage{array}
\usepackage[T1]{fontenc} % for special characters in some author names

\newcommand{\xraychap}{X-ray Chapter}

\newcommand{\optchap}{Optical Chapter}
\newcommand{\radiochap}{Radio Chapter}
\newcommand{\ehcochap}{Echo Chapter}
\newcommand{\wdchap}{White Dwarf Chapter}
\newcommand{\impostchap}{Imposters Chapter}

\newcommand{\disrupchap}{Disruption Chapter}
\newcommand{\flowchap}{Formation of the Accretion Flow Chapter} % shorter name? Circulation perhaps? :)
\newcommand{\diskchap}{Accretion Disc Chapter}

 \journalname{ISSI Book on TDEs}

\usepackage{ref_macros} 
\begin{document}

\title{Radiative Emission Mechanisms}
%\subtitle{Subtitle}


\author{Nathaniel Roth \and
        Elena Maria Rossi \and
        Julian Krolik \and 
        Tsvi Piran \and 
        Brenna Mockler \and
        Daniel Kasen
        %etc.
}


\institute{Nathaniel Roth \at
              Joint Space Science Institute, University of Maryland, College Park, MD 20742, USA  \\
              \email{nroth@astro.umd.edu} %//
           \and
           Elena M. Rossi \at
           Leiden Observatory, Leiden University, PO Box 9513, 2300 RA, Leiden, the Netherlands
           \and
           Julian H. Krolik \at
           Dept. of Physics and Astronomy, Johns Hopkins University, Baltimore, MD 21218, USA 
           \and
           Tsvi Piran \at 
           Racah Institute of Physics, Hebrew University, Jerusalem 91904, Israel
           \and
           Brenna Mockler \at
           Dept. of Astronomy and Astrophysics, University of California, Santa Cruz, CA 95064, USA
           \and
           Daniel Kasen \at
           Depts. of Physics and Astronomy, University of California, Berkeley, Berkeley, CA 94720 USA
}


\maketitle

\begin{abstract}
We describe how the various outcomes of stellar tidal disruption give rise to observable radiation. We separately consider the cases where gas circularizes rapidly into an accretion disc, as well as the case when shocked debris streams provide the observable emission without having fully circularized. For the rapid circularization case, we describe how outflows, absorption by reprocessing layers, and Comptonization can cause the observed radiation to depart from that of a bare disc, possibly giving rise to the observed optical/UV emission along with soft X-rays from the disc. If, instead, most of the debris follows highly eccentric orbits for a significant time, many properties of the observed optical/UV emission can be explained by the scale of those eccentric orbits and the shocks embedded in the debris flow near orbital apocenter.  In this picture, soft X-ray emission at early times results from the smaller amount of debris mass deflected into a compact accretion disc by weak shocks near the stellar pericenter. A general proposal for the near-constancy of the ultraviolet/optical color temperatures is provided, by linking it to incomplete thermalization of radiation in the atmosphere of the emitting region. We also briefly discuss the radio signals from the interaction of  unbound debris and jets with the black hole environment.
%Include keywords.

% \PACS{PACS code1 \and PACS code2 \and more} % not needed? 
% \subclass{MSC code1 \and MSC code2 \and more} %not needed, mathematical subject classification 
\end{abstract}

\newpage
\section{Introduction}

In this chapter we describe how the various components of the disrupted star emit the radiation we observe. We immediately face the difficulty that, despite much progress, the underlying hydrodynamics are not fully understood and remain the subject of active research (see for example the \flowchap\ and \diskchap). We also expect the disruption outcome to vary substantially among separate events. We therefore describe the emission that might result from a variety of post-disruption scenarios, with the understanding that no single picture is capable of explaining every feature of the ever-growing observational dataset. 

A summary of the TDE components which may give rise to detectable radiation is provided in Table~\ref{tab:EmissionTable}. As indicated, a subset of these are discussed in detail in this chapter. For the others, we have referred readers to relevant  chapters in this review collection, or to specific papers. 
\begin{table*}
\begin{tabular}{ |m{12em}|m{16mm}|m{18em}|}
 \multicolumn{3}{|c|}{\textbf{Components of TDEs posited to give rise to radiative emission}} \\
 \hline
 \hline
 Emission source & Primary detection bands & Locations discussed \\
 \hline
 \hline
 Flash from vertical compression shock during first passage  & Soft X-ray, Hard X-ray, $\gamma$-ray & \citet{Kobayashi:2004a,Brassart:2008a,Brassart:2010a,Guillochon:2009a,Stone:2013a,Yalinewich2019-1,Gafton2019}, and \disrupchap\\
 \hline
 Nuclear burning and/or detonation following vertical compression at pericenter & Optical, UV  & \emph{Main sequence stars:} \citet{Carter:1982a,Carter:1983a,Bicknell:1983a,Pichon1985,Luminet:1989a,Brassart:2008a}. \emph{White dwarfs:} \wdchap \\
 \hline
 Electron recombination and cooling in unbound debris  & Optical, UV & \citet{Kochanek:1994a, Kasen:2010a}\\
 \hline
 Accretion flow and associated thermal or quasi-thermal emission & Soft X-ray, UV & Section~\ref{sec:BareDiskAllTimes} and references therein, \diskchap, \xraychap \\
 \hline
 Outflow launched from location of first stream collision & Optical, UV & Section~\ref{sec:OutflowsAdiabatic}, \flowchap\\
 \hline
 Disc outflow launched by radiation pressure & Optical, UV, Soft X-ray &  Section~\ref{sec:OutflowsAdiabatic}, section~\ref{sec:Mosfit}, \diskchap\\
 \hline
Reprocessing of inner accretion emission by material farther out (e.g. thick disc, outflowing disc material, or incompletely circularized bound material) & Optical, UV & Section \ref{sec:ReprocessingRequirements}, section \ref{sec:Mosfit}, section \ref{sec:PartialThermFreefree}, \optchap \\ 
 \hline
 Shock-heated, incompletely circularized stellar debris & Optical, UV & Section~\ref{sec:ShockedStreamsAll}, \flowchap, \optchap \\ 
 \hline
Inverse Comptonization of photons by thermal and/or non-thermal electrons in a jet & Soft X-ray, Hard X-ray, $\gamma$-ray & Section~\ref{sec:EarlyDiskEvolution}, Section \ref{sec:Comptonization}, section \ref{sec:JetRadio}, \xraychap \\
 \hline
 Forward shock driven by jets encountering circum-nuclear material & Radio & Section~\ref{sec:JetRadio}, \radiochap \\
 \hline
 Mildly relativistic outflow interacting with circum-nuclear material  & Radio & Section~\ref{sec:UnboundDebris}, \radiochap\\
 \hline
 Unbound debris interacting with circum-nuclear material & Radio & Section~\ref{sec:UnboundDebris}, \radiochap \\
 \hline
 Dust heated by UV and X-rays from TDE & IR & Section~\ref{sec:ReprocessingAllTypes}, \ehcochap \\
 \hline
 \end{tabular}
 \label{tab:EmissionTable}
 \caption{A listing of sources of radiative emission from tidal disruption events by super-massive black holes.}
\end{table*}

In section \ref{sec:RapidCirc} we summarize the emission from accretion discs (and their outflows) that may form from TDEs, which are continuously fed and potentially super-Eddington at the outset. In section \ref{sec:ShockedStreamsAll} we discuss the emission from shock-heated and elliptically orbiting tidal debris. In section \ref{sec:LinesAndContinuum} we consider general principles governing optical and ultraviolet (UV) line and continuum emission, which may apply to several of the above scenarios. Finally, section \ref{sec:BhEnvironsAll} addresses the radio emission from the interaction of unbound stellar debris and/or accretion disc jets with the surrounding circum-nuclear environment.

\section{Emission from accretion discs and their outflows, for material that rapidly circularizes }
\label{sec:RapidCirc}
In this section we will consider the radiative emission that is produced if a large fraction of the bound stellar material is capable of rapidly dissipating its orbital energy and entering a circularized accretion flow around the black hole. In section~\ref{sec:EarlyDiskEvolution} we first consider the classical picture for the observational electromagnetic properties, dating back to \citet{Rees:1988a} and \citet{Phinney1989}, where thermal emission peaking in soft X-rays is emitted by a circularised accretion disc fed by the continuous stream of stellar debris.  We will track the emission properties of such an unobscured accretion disc with an accretion rate that is initially tied to the post-disruption mass fallback rate. In section~\ref{sec:LateDiskEvolution} we examine how the accretion flow is expected to behave several years after disruption, when the accretion rate is governed by viscous processes within the disc, rather than the mass fallback rate. While late-time disc emission models have been successfully matched to UV and X-ray data \citep{vanVelzen2019, Jonker2019}, the bare discs of section~\ref{sec:EarlyDiskEvolution} generally do not fit observational data well on their own for the first few years after detection by optical surveys. In particular, the bare disc models often over-predict the amount of soft x-rays emitted when compared to observed limits, and vastly under-predict the observed optical and UV emission at early times.  Thus, in section~\ref{sec:ReprocessingAllTypes} we will describe a number of factors that may cause the observed radiation to depart from the bare disc at early times, which may shift the peak of the emitted spectrum away from the soft X-rays, towards UV and optical wavelengths. These include adiabatic reprocessing of radiation in outflows (section~\ref{sec:OutflowsAdiabatic}); absorption, re-thermalization and re-emission of inner disc emission by material farther out (section~\ref{sec:ReprocessingRequirements}); and Comptonization (section~\ref{sec:Comptonization}). In section~\ref{sec:Mosfit}, we discuss a model for fitting multi-band TDE light curves based on tying the light curves to the post-disruption mass fallback rate and allowing for a dynamically adjusting photosphere.

\subsection{Bare disc emission}
\label{sec:BareDiskAllTimes}
%authors: EMR and NR
\subsubsection{Disc properties at early times ($<$ a few years after disruption)}
\label{sec:EarlyDiskEvolution}
The prototype system we adopt consists of a main sequence star with mass $M_\ast$ and radius $R_\ast$ disrupted on an orbit with pericenter radius $r_{\rm p}$ equal to the tidal radius $r_{\rm t} \equiv R_\ast \left(M_{\rm BH} / M_{\ast} \right)^{1/3} $ where $M_{\rm BH}$ is the mass of the black hole (i.e. we set the so-called ``penetration factor" $\beta \equiv r_{\rm p}/r_{\rm t}$ to 1 where $r_p$ is the pericenter distance from the black hole of the initial stellar orbit) . For scaling relations we will adopt the short-hand $M_{\rm BH,6} \equiv M_{\rm BH} / ( 10^6 M_\odot )$. We neglect the role of black hole spin unless stated otherwise.

After stellar disruption, if self-gravity in the stellar debris can be ignored, then the debris move ballistically, i.e. under the sole influence of the black hole's gravity. The portions of the debris that remain bound follow highly eccentric ($1-e \leq 0.02$) elliptical orbits returning to the site of the star's disruption at a time-varying rate. We let $t = 0$ refer to the time of pericenter passage of the initial stellar orbit. Assuming for the moment a flat distribution of orbital energy for the stellar debris, the ballistic dynamics implies a fallback rate of
\begin{equation}
\label{MassFallbackTophat}
\dot{M}_{\rm fb} = \frac{1}{3} (M_*/t_{\rm fb}) (t/t_{\rm fb})^{-5/3} \qquad {\rm for} \; t \ge t_{\rm fb} \, \, , \,\, 0 \,\, {\rm otherwise}
\end{equation}
where $t_{\rm fb}$ is the orbital period of the most bound debris (thus the first to come back to the tidal radius), and evaluates to
\begin{equation}
t_{\rm fb} \approx 40 \,\, {\rm days} \,\, (M_{\rm BH,6})^{1/2} \left(\frac{M_{\ast}} {M_\odot} \right)^{-1}\left(\frac{R_\ast}{R_\odot} \right)^{3/2}.
% 41 to 2 sig figs
\end{equation}
This approximation for $\dot{M}_{\rm fb}$ is taken for simplicity. and the implications of this simplification will also be discussed below.

In this scenario, the timescale for the stellar debris to lose the necessary energy to circularize is assumed to be short compared to $t_{\rm fb}$ (see section \ref{sec:ShockedStreamsAll} for challenges to this assumption). Within the newly formed disc, most of the mass is initially deposited at a distance $r_{d} \approx 2 r_{t}$, as demanded by the conservation of the star's specific angular momentum $j^2 = (2GM_{\rm BH} r_{t})$ on its originally parabolic orbit with pericenter $r_{t}$ where $G$ is Netwon's gravitational constant. This evaluates to \begin{align} 
r_{d} &\approx 100 \, r_{g} \left( M_{\rm BH,6} \right)^{-2/3} \left ( \frac{M_{\ast}} {M_\odot} \right)^{-1/3}\left(\frac{R_\ast}{R_\odot} \right) \nonumber \\
&\approx 10^{13} \,\,{\rm cm} \, \, \left( M_{\rm BH,6} \right)^{1/3} \left ( \frac{M_{\ast}} {M_\odot} \right)^{-1/3}\left(\frac{R_\ast}{R_\odot} \right),
%94, 1.4e13 to 2 sig figs
\end{align}
 where $r_g \equiv G M_{\rm BH}/c^2$. Meanwhile, the high rate of mass accretion through the disc may lead it to be geometrically thick, with a vertical scale height to radius ratio $h/r \sim 1$ (we will elaborate on this point shortly). In this case the viscous (disc inflow) timescale $t_{\nu}$ at $r_d$ is initially much shorter than the fallback timescale:
\begin{align}
\frac{t_{\nu}(r_d)}{t_{\rm fb}} &\approx \alpha^{-1} \left(\frac{h}{r}\right)^{-2} \frac{P(r_{d})}{2 \pi t_{\rm fb}} \nonumber \\ &\approx 10^{-2} \left(\frac{\alpha}{0.1}\right)^{-1} \left(\frac{h}{r}\right)^{-2} \left( M_{\rm BH,6} \right)^{-1/2}  \left(\frac{M_\ast}{M_\odot}\right)^{1/2},
\label{eq:tnutmin}
%1.3e-2 to 2 sig figs
\end{align}
where $P(r_{\rm d})\approx 8$ hours $\ll t_{\rm fb}$ is the orbital period at $r_d$ and $\alpha$ is the standard Shakura \& Sunyaev viscosity parameter \citep{Shakura1973} with a value appropriate for thick discs. As a result of this timescale ordering, material is immediately accreted through the disc. The rate at which mass feeds the black hole, $\dot{M}_{\rm acc}$, is then given by 
\begin{equation}
\dot{M}_{\rm acc} = f_{\rm in}\dot{M}_{\rm fb} \, \, ,
\label{eq:mdotacc}
\end{equation}
where the factor $f_{\rm in} \le 1$ accounts for mass launched in an outflow during circularization and/or accumulating in the disc. Generally we will expect $f_{\rm in}$ to be a function of time; however, for the purpose of deriving an approximate time dependence for $\dot{M}_{\rm acc}$ we will take $f_{\rm in}$  to be constant with time so that $\dot{M}_{\rm acc} \propto \dot{M}_{\rm fb}$.

Accretion through the disc will cause orbital energy to be dissipated as heat, and some fraction of this dissipated energy will escape as radiation. We will label the radially-integrated bolometric luminosity of this radiation, with contributions from both sides of the disc, as $L_{\rm bol}$, and we will often make reference to the instantaneous radiative efficiency $\eta = L_{\rm bol}/(\dot{M}_{\rm acc} c^2)$ as a function of time.

Notably, the fallback rate implies accretion rates that may exceed a critical value defined as the rate that would lead to the Eddington luminosity $L_{\rm Edd}$ if the radiative efficiency were set to a constant reference value $\eta_{\rm ref}$:
\begin{equation}
\dot{M}_{\rm cr} \equiv  \frac{L_{\rm Edd}}{\eta_{\rm ref} c^2}.
\label{m_acc}
\end{equation}
We will be using the definition $L_{\rm Edd} \equiv  4 \pi G M_{\rm BH}\,c / \kappa_{\rm es}$ where $\kappa_{\rm es}$ is the electron scattering opacity.  It will be useful to consider the time-dependent ratio $f_{\rm Edd}$ of the instantaneous value of $\dot{M}_{\rm acc}$ over $\dot{M}_{\rm cr}$:
\begin{align}
f_{\rm Edd} &\equiv \frac{\dot{M}_{\rm acc}}{\dot{M}_{\rm cr}}  =  \frac{f_{\rm in}\, \dot{M}_{\rm fb}\, \eta_{\rm ref} {c^2}}{L_{\rm Edd}} \nonumber \\ &\approx 100 \, f_{\rm in}  \left(\frac{\eta_{\rm ref}}{0.1}\right) \left(\frac{\kappa_{\rm es}}{0.34 \, {\rm cm}^2 \, {\rm g}^{-1}}\right) \nonumber \\ &\times \left(M_{\rm BH,6} \right)^{-3/2} \left( \frac{M_\ast}{M_\odot} \right)^2 \left( \frac{R_\ast}{R_\odot} \right)^{-3/2}  \left(\frac{t}{t_{\rm fb}}\right)^{-5/3}.
\label{f_edd}
%110 to 2 sig figs
\end{align}
We can use equation~\eqref{f_edd} to provide an estimate of the time after disruption $t_{\rm cr}$ when the mass fallback rate transitions from  super-critical to sub-critical, which is when $f_{\rm Edd}$ drops to 1:
\begin{align}
t_{\rm cr} &\approx 700 \,\,{\rm days} \, \left(f_{\rm in}\right)^{3/5} \left(\frac{\eta_{\rm ref}}{0.1} \right)^{3/5} \, \left(\frac{\kappa_{\rm es}}{0.34 \, {\rm cm}^2 \, {\rm g}^{-1}}\right)^{3/5} \nonumber \\
&\times \left(M_{\rm BH,6} \right)^{-2/5} \left( \frac{M_\ast}{M_\odot} \right)^{1/5}  \left( \frac{R_\ast}{R_\odot} \right)^{3/5}\,\, .
%7.0e2 days to 2 sig figs
\label{t_cr}
\end{align}
The time-dependence in equation~\eqref{f_edd} follows from the approximations stated at the beginning of this section, in addition to the assumption that $f_{\rm in}$ does not vary with time. 
The power-law time-dependence for $\dot{M}_{\rm fb}$ is closest to reality after a few $t_{\rm fb}$ from the return of the most bound debris, and the actual power-law may deviate from $-5/3$ (see the \disrupchap). At earlier times the star's structure imprints a characteristic rising-to-a-peak shape, with some dependence on the impact parameter of the initial stellar orbit. Hydrodynamic simulations show that the peak $\dot{M}_{\rm fb}$ remains super-critical for our fiducial parameters but it is suppressed by a factor of a few compared to the peak value stated in equation~\eqref{MassFallbackTophat} \citep[e.g][]{Evans:1989a}. Interestingly, since $\dot{M}_{\rm fb}/\dot{M}_{\rm cr} \propto M_{\rm bh}^{-3/2}$, only black holes with masses below $ \sim 10^{7} M_{\odot}$ may present these super-critical mass accretion flows (neglecting the effects of black hole spin). For other considerations concerning how the transition between super- and sub-critical accretion flow depends on $M_{\rm BH}$, see \citet{Wu2018}. We also note that the super-criticality of $\dot{M}_{\rm acc}$ as we have discussed it here depends crucially on $f_{\rm in}$ not being too small, and on our initial assumption of rapid circularization. 

Super-critical accretion is a topic of ongoing research, and the most recent results on such accretion flows in the context of TDEs is presented in the \diskchap. Here we summarize some analytic results that are at least qualitatively in line with more detailed numerical calculations. At the highest accretion rates ($f_{\rm Edd} \gtrsim 100$), the energy transport through the disc is dominated by advection of trapped radiation, as the large optical depth of the disc leads to the radiative cooling timescale is much longer than the timescale that trapped radiation is advected inward \citep[e.g.][]{Begelman1978,Paczynsky1980,Narayan1994}. This trapping of radiative energy leads to a large disc scale height, justifying our earlier decision to set $h/r \sim 1$ in equation~\eqref{eq:tnutmin}. For more near-critical accretion rates ($0.1 \lesssim f_{\rm Edd} \lesssim 100$) the trapping of radiation is reduced yet still important. Potential complications in this regime are thermal and viscous instabilities \citep{Shakura1976, Lightman1974}. Some of the consequences of these instabilities for accretion following tidal disruption were investigated by \citet{Shen:2014b}.  However, recent global radiation magneto-hydrodynamic simulations of sub-critical ($f_{\rm Edd} \simeq 0.1$) accretion discs around super-massive black holes have found that, for certain magnetic field configurations, these instabilities are not triggered \citep{Jiang2019-2}. Another important consideration is that outflows are generally expected in advection-dominated flows, as discussed in the literature concerning advection-dominated inflow-outflow solutions \citep{Blandford1999, Blandford2004} and in numerical calculations \citep[e.g.][]{Sadowski2014, Jiang2019}. We will have more to say about the role of outflows in section \ref{sec:OutflowsAdiabatic}. 

The near-critical regime is of the greatest interest to us. Putting aside the issue of stability, we may make use of a formula for the steady-state effective temperature (a measure of the radiative flux) of disc annuli as a function of radius and $\dot{M}_{\rm acc}$ provided by \citet{Strubbe:2009a} (henceforth SQ09):
\begin{equation}
\label{SlimDiskTemperatures}
\sigma_{\rm SB} T_{\rm eff}^4 = \frac{3 G M_{\rm BH} \dot{M}_{\rm acc}f_{\rm nt}}{8 \pi r^3} \left[\frac{1}{2} + \left\{ \frac{1}{4}  + 6 f_{\rm nt} \left( \frac{\dot{M}_{\rm acc}c^2}{L_{\rm Edd}}\right)^2\left(\frac{r}{r_g} \right)^{-2} \right\}^{1/2} \right]^{-1},
\end{equation}
where $\sigma_{\rm SB}$ is the Stefan-Boltzmann constant and $f_{\rm nt} \equiv 1 - \sqrt{r_{\rm ISCO}/r}$, which comes from imposing a no-torque boundary condition at the innermost stable circular orbit (ISCO) \citep[this may be compared to more recent work by][who allow for finite ISCO stress and a fully relativistic, time-dependent treatment]{Mummery2020}. Note that $\dot{M}_{\rm acc}c^2/L_{\rm Edd}$ is equal to $f_{\rm Edd} / \eta_{\rm ref} $ as we have defined it. The temperature profile in equation  \eqref{SlimDiskTemperatures} becomes identical to the thin disc result of \citet{Shakura1973} for $f_{\rm Edd} \ll 1$. The assumptions underlying equation~\eqref{SlimDiskTemperatures} are related to the so-called ``slim disc'' models of \citet{Abramowicz1988} which apply to discs with accretion rates that are close to the critical value, although in deriving the SQ09 formula, general relativistic modifications to the black hole's gravitational potential are ignored when computing the orbital velocities of the gas. 

Equation~\eqref{SlimDiskTemperatures} is a simplified version of what would be computed from a detailed hydrodynamic simulation of the accretion process, which would compute physical stresses driving the accretion and account for cooling using radiative transfer.  We provide this formula to give a concrete example of how $L_{\rm bol}$ and $\eta$ can be computed once $T_{\rm eff}(r)$ has been determined: we integrate the flux over disc annuli, accounting for both sides of the disc (again ignoring relativistic effects):
\begin{equation}
    \label{eq:LbolDiskIntegral}
    L_{\rm bol} = 2 \int \sigma_{\rm SB} T_{\rm eff}^4(r) 2 \pi r dr
\end{equation}

For $f_{\rm Edd} \ll 1$, then for this simplified disc model, as for \citet{Shakura1973}, $L_{\rm bol}$ for a disc of a given size varies linearly with $\dot{M}_{\rm acc}$. In other words, $\eta$ is constant in this regime for fixed disc size. On the other hand, for $f_{\rm Edd} \gg 1 $, equation~\eqref{SlimDiskTemperatures} indicates that the radiative efficiency drops with increasing $\dot{M}_{\rm acc}$, so that $L_{\rm bol}$ varies sub-linearly with $\dot{M}_{\rm acc}$. This is the result of increased trapping of radiation at larger values of $f_{\rm Edd}$. As a consequence, while the total radiative luminosity of the disc may exceed $L_{\rm Edd}$, the rate at which the luminosity grows beyond this value is limited. For rough approximations, a common practice is to allow $L_{\rm bol}$ to grow with $\dot{M}$ according to the thin-disc result until $L_{\rm Edd}$ is reached, and then to cap $L_{\rm bol}$ at $L_{\rm Edd}$ for larger values of $\dot{M}_{\rm acc}$. This approximation, in addition to neglecting the more detailed growth of $L_{\rm bol}$ near and above $L_{\rm Edd}$, also neglects how the change in the disc size with time will affect $L_{\rm bol}$, in addition to a changing $\dot{M}_{\rm acc}$.

In order to get closer to observations, the luminosity in a given band, rather than the bolometric one, needs to be derived. If the radiation can effectively thermalize in the disc atmosphere, the disc radiates as a superposition of blackbody spectra from disc annuli, with a temperature profile equal to the effective temperature profile specified in equation~\eqref{SlimDiskTemperatures}, with generally outwards-decreasing peak frequency and flux. However, if the ionization state of the disc becomes high enough, electron scattering opacity is expected to be comparable to or exceed absorption opacities that thermalize the radiation, preventing local thermodynamic equilibrium (LTE) from being established in the disc atmosphere. This may lead to several departures from the full thermalization prediction, the simplest of which is a color correction $f_{\rm col}(r)$ that measures the degree to which the local color temperature of the radiation exceeds the local effective temperature. While such color corrections are most often applied to X-ray binary accretion discs \citep{Shakura1973,Shimura1995, Davis2005}, recent work also indicates they may apply to some accreting super-massive black holes such as those of narrow-line Seyfert 1 galaxies, with values of $f_{\rm col}$ reaching 2 -- 3 at small radii \citep{Done2012}. We will have more to say about non-LTE effects on the emission observed from TDEs, including emitting components other than the disc, in sections \ref{sec:ReprocessingRequirements} and \ref{sec:PartialThermFreefree}. For now, we simply incorporate these effects on the disc continuum emission by implicitly allowing for a color correction whenever a temperature is specified below.

%figure slim disc Eddington limited + wind à la Lodato and Rossi 2011
\begin{figure}[]
\centering
\includegraphics[scale=0.6]{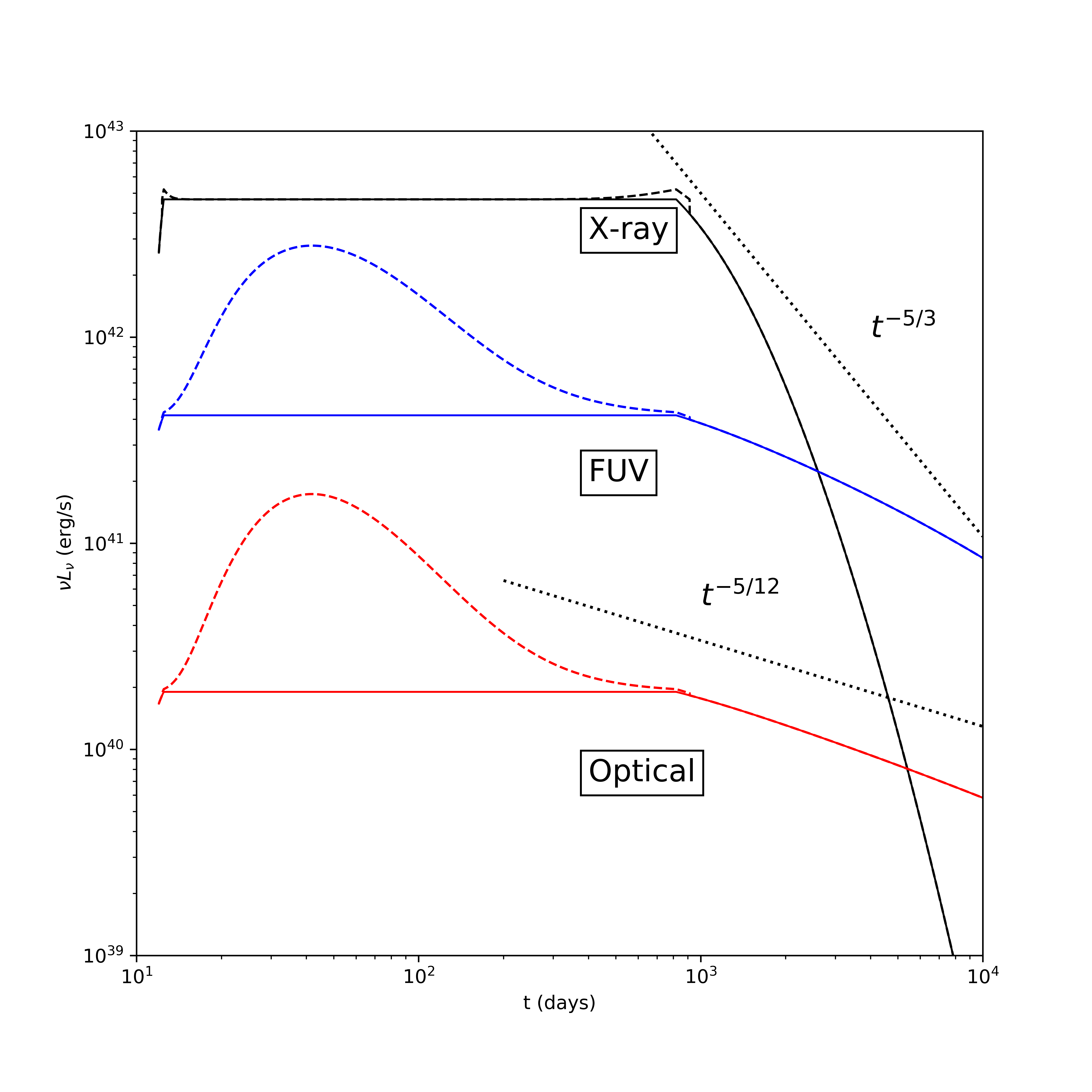}
\caption{Specific luminosity in soft X-rays, far ultraviolet (FUV) and optical radiation as a function of time for an accretion disc-plus-outflow as discussed in sections~\ref{sec:EarlyDiskEvolution} and \ref{sec:OutflowsAdiabatic}. Solid lines are light curves from an analytic model of a super-critical , Eddington limited accretion disc (equation~\ref{SlimDiskTemperatures}) with $\dot{M}_{\rm acc}$ proportional to the fallback $\dot{M}$, while the dashed lines include the emission from a wind component, as described in \cite{Lodato:2011a}. We here consider a $10^{6} M_{\odot}$ black hole, a solar mass star, a pericenter equal the tidal radius, and we assume that 10\% of the accretion rate is lost from the disc into the wind.}
\label{fig:disc_mw_emission}
\end{figure}

Following \citet{Miller:2015a}, it is useful to determine the maximum value of $T_{\rm eff}$ that is reached in the hottest disc annulus, as a function of $\dot{M}_{\rm acc}$. Using equation~\eqref{SlimDiskTemperatures}, we find that near the ISCO, 
\begin{eqnarray}
T_{\rm max} &\approx& 4.3 \times 10^5 \,  {\rm K} \left( M_{\rm BH,6}\right)^{-1/4} \left( \frac{\kappa_{\rm es}}{0.34 \, \, {\rm cm}^2 \, {\rm g}^{-1} }\right)^{-1/4} \nonumber \\ &{\rm for}& \,\, f_{\rm Edd} \gg 1, \nonumber \\ &\approx&  2.4 \times 10^5 \,  {\rm K} \left( \frac{f_{\rm Edd}}{0.1}\right)^{1/4} \left( M_{\rm BH,6}\right)^{-1/4} \nonumber \\&\times& \left(\frac{\eta_{\rm ref}}{0.1}\right)^{-1/4} \left( \frac{\kappa_{\rm es}}{0.34 \, {\rm cm}^2 {\rm g}^{-1} }\right)^{-1/4} \nonumber \\  &{\rm for}& \,\, f_{\rm Edd} \ll 1.
\end{eqnarray}
Allowing for finite fluid stress at the ISCO can lead to higher maximum disc temperatures in the case of sub-critical accretion onto a black hole of a given mass \citep{Mummery2020}. Note that for $f_{\rm edd} \gg 1$, the maximum disc temperature approaches a value that it is independent of $f_{\rm edd}$. Meanwhile, an approximate scaling for $T_{\rm eff}$ of the outer disc, which we call $T_{\rm out}$, based on equation \eqref{SlimDiskTemperatures} for $r_d/r_g \gg 1$, is
\begin{eqnarray}
T_{\rm out} &\approx& 1.6 \times 10^5 \,  {\rm K} \left( M_{\rm BH,6}\right)^{-1/4} \left( \frac{\kappa_{\rm es}}{0.34 \, \, {\rm cm}^2 \, {\rm g}^{-1} }\right)^{-1/4}  \left( \frac{r_d}{100 \, r_g} \right)^{-1/2} \nonumber \\ &{\rm when}& \,\, \frac{f_{\rm Edd}}{\eta_{\rm ref}}\frac{r_g}{r_d} \gg 1, \nonumber \\ &\approx&  6.1 \times 10^4 \,  {\rm K} \left( \frac{f_{\rm Edd}}{0.1}\right)^{1/4} \left( M_{\rm BH,6}\right)^{-1/4} \nonumber \\&\times& \left(\frac{\eta_{\rm ref}}{0.1}\right)^{-1/4} \left( \frac{\kappa_{\rm es}}{0.34 \, {\rm cm}^2 {\rm g}^{-1} }\right)^{-1/4} \left( \frac{r_d}{100 \, r_g} \right)^{-3/4} \nonumber \\ &{\rm when}& \,\, \frac{f_{\rm Edd}}{\eta_{\rm ref}}\frac{r_g}{r_d} \ll 1.
\end{eqnarray}
These temperatures are high enough that the optical luminosity mainly comes from the outer radii of the disc and sits in the Rayleigh-Jeans portion of the spectrum for emission at $T_{\rm out}$. In the case $(f_{\rm Edd}/{\eta})(r_{\rm g}/r_{\rm d}) \gg 1 $, the optical luminosity scales roughly as $\nu L_{\nu,\rm opt} \propto T_{\rm out}$, which depends only on how far the disc extends. Once the mass accretion rate has fallen low enough so that $(f_{\rm Edd}/{\eta})(r_{\rm g}/r_{\rm d}) \ll 1 $  we then have $T _{\rm out} \propto \dot{M}_{\rm acc}^{1/4}$, and so for the approximate scaling $ \dot{M}_{\rm acc} \propto t^{-5/3}$ we arrive at  $\nu L_{\nu,\rm opt} \propto t^{-5/12}$ \citep[see Fig.\ref{fig:disc_mw_emission}, and][]{Strubbe:2009a, Lodato:2011a}. The duration of this temporal behaviour at optical wavelengths depends on several factors, such as whether $t_\nu / t_{\rm fb}$ at $r_d$ approaches unity (see equation~\ref{eq:tnutmin}), or the black hole accretion state changes in other ways at lower $\dot{M}_{\rm acc}$, or whether the TDE emission is outshone by that of the host galaxy. 

On the other hand, observed soft X-ray frequencies (photon energies $\gtrsim$ 0.2 keV) may initially fall close to the peak of the disc emission spectrum, in which case the X-ray light curve should more closely track the bolometric luminosity. As previously indicated, the bolometric luminosity may be capped near Eddington initially, and later decline as $\nu L_{\nu, \rm X} \propto t^{n}$ ($n \approx -5/3$). As shown in Fig.~\ref{fig:disc_mw_emission}, for a $10^{6} M_{\odot}$ black hole, this power-law phase -- following the Eddington limited stage --  precedes an exponential decline when the spectrum softens to the point that the observed soft X-ray band overlaps only with the Wien tail of the spectrum emitted by the innermost parts of the disc \citep{Lodato:2011a}. For yet another possibility, \citet{Mummery2020} considered the case where the soft X-ray band is initially located on the Wien tail of the spectrum for an initially \emph{sub-critical} disc, and the disc inflow time remains non-negligible compared to the duration of the observations. They found that the X-ray light curve is then described by a power-law multiplied by an exponential, with the power-law being tied to the viscous evolution of the disc, rather than the mass fallback rate (see section~\ref{sec:LateDiskEvolution}). We noted that \citet{Mummery2020} initialized their disc as a ring of material at radius 15 $r_g$ from the black hole, which if interpreted as the initial circularization radius, corresponds to a very close pericenter passage for the star ($r_p = 7.5 r_g$).

Observationally, some X-ray TDE candidates seem to show an approximate $t^{-5/3}$ decline in the soft X-ray band (\xraychap), which is consistent with the disc models discussed in this chapter during the \emph{sub-critical} mass inflow phase, for both  the luminosity and temperature. The X-ray light curve of the event ASASSN-14li \citep{Jose2014,Holoien:2016b} can be fit by the bare disc model of \citet{Mummery2020}, which was also sub-critical. In contrast, the event discussed in \citet{Lin:2017a} does suggest an Eddington-limited plateau in the early X-ray emission. Aside from that event, however, there is little evidence for sustained, Eddington-limited phases in the X-ray data lasting as long as predicted by equation \ref{t_cr} with $f_{\rm in} \sim 1$. Moreover, optically selected TDEs have constraining X-ray upper limits, at least in the first few months following detection, and their optical specific luminosities are at least an order of magnitude larger than predicted from bare disc models. This can be interpreted in several ways. In addition to explanations for this fact described in this chapter (sections \ref{sec:ReprocessingAllTypes} and \ref{sec:ShockedStreamsAll}), another possibility is that the low temperatures of TDE flares result from line-driven winds from an extended accretion disc \citep[along the lines of][]{Laor2014}, which suppresses the fraction of the accreting gas which reaches the hotter inner disc \citep{Miller:2015a}. In this case, in order to explain the stringent X-ray upper limits (sometimes $L_X < 10^{41}$ erg s$^{-1}$), the fraction of gas that reaches the inner disc would need to be as low as $10^{-4}$.

We leave this section by noting an additional puzzle when comparing X-ray TDEs to steadily accreting super-massive black holes in active galactic nuclei (AGN). With the exception of the rare TDEs that have produced relativistic jets, the early X-ray spectra of TDEs are systematically ``softer'', i.e. weighted toward lower photon energies, than a broad sample of flaring AGN. Early TDE spectra also do not harden with decreasing flare luminosity, as is typically the case for AGN flares \citep{Auchettl2018}. This soft emission is indeed expected for the thermal emission from a disc as described above. However, AGN spectra typically also possess an emission component of non-thermal, higher energy photons, which are attributed to a hot ``corona'' which may produce high-energy emission through inverse Compton scattering \citep[e.g.][]{Haardt1991}. Some unspecified mechanism seems to inhibit the growth of such coronae during the first few years of TDE accretion disc evolution. Some authors \citep[e.g.][]{Wevers2019} have suggested a connection between the soft X-ray spectra of TDEs in their first few years following disruption and the X-ray properties of the AGN class known as narrow-line Seyfert 1 galaxies, which have systematically higher accretion rates (in the sense of $f_{\rm Edd}$) and systematically softer X-ray spectra than the AGN population at large. However, there are limits to how far this analogy can be taken (see the \impostchap\ for more details). 

\subsubsection{Disc emission at later times (at least several years after disruption)}
\label{sec:LateDiskEvolution}
After $t_{\rm cr}$, as the accretion rate has decreased and dropped below $\dot{M}_{\rm cr}$, a state transition takes place when the thick disc we have postulated becomes geometrically thin ($h/r \ll 1$). As a consequence, the viscous timescale for the outer disc increases and eventually becomes longer than the fallback timescale at the outer edge of the disc (equation~\ref{eq:tnutmin}), where material increasingly accumulates. If the disc remains viscously and thermally stable, the expectation is that internal viscous processes now determine the mass accretion flow, flattening the bolometric light curve to $L_{\rm bol} \propto \dot{M}_{\rm acc} \propto t^{-1.2}$ when the disc evolution is calculated in the Newtonian limit for the gas orbital motion and using a no-torque inner boundary condition \citep{Cannizzo:1990a,Shen:2014b}. Allowing for finite stress at the ISCO, and analyzing the thin disc equations using full general relativity, yields a shallower asymptotic decline, $L_{\rm bol} \propto t^{-n}$ with $n \approx 0.6$ - $0.7$ \citep{Balbus2018}, and X-ray emission on the Wien tail then continues to follow the power-law-times-exponential decline discussed earlier \citep{Mummery2020}. The predicted light curve for frequency bands close to the Rayleigh-Jeans part of the spectrum is  $L_{\nu} \propto \dot{M}_{\rm acc}^{1/4} \propto t^{-0.3}$ for the Newtonian, vanishing-inner-stress predictions, and $\approx t^{-0.16}$ for the updated treatment by \citet{Balbus2018}. In all cases, the viscous spreading of the disc increases the emitting area of the outer disc annuli and causes the actual Rayleigh-Jeans band light curves to decline even more gradually before they approach the asymptotic power-law decays quoted here.

Indeed, slow (nearly-flat) temporal decay in all three bands of the Swift UVOT instrument is seen in in a sample of TDEs followed up 5--10 years after first detection \citep{vanVelzen2019}. The characteristic photospheric radii are close to the tidal disruption radii, as expected for discs that have spread out by just a factor of a few. Late-time X-ray detections of TDEs, particularly those without early X-ray detections, also support the interpretation that a bare accretion disc is responsible for the late-time emission \citep{Jonker2019}. Interestingly, the spectral energy distribution of the disc model that can explain the late-time observations suggest that most of the radiation is emitted in the hardly observable extreme-ultraviolet frequency band. This is consistent with some observed TDE dust echoes (\ehcochap). The energy radiated over time is estimated to be a few percent of the available accretion energy, a cumulative radiative efficiency that is not much lower than many AGN, providing a possible solution to one aspect of what was dubbed ``the missing energy problem'' prior to these late-time observations \citep{Piran:2015b, Stone:2016a, Lu2018-1}.

\subsection{Reprocessing by outflows, radiative absorption, and Comptonization that affect the early UV/optical emission}
\label{sec:ReprocessingAllTypes}

As discussed elsewhere in this book (e.g. the \optchap), many TDEs exhibit optical and UV emission at early times (weeks to months after first detection) far in excess of what is predicted from the preceding description of bare disc emission. Similarly, their x-ray emission is, at least initially, much lower than the bare disc predictions for the case of rapid inflow and circularization. Thus, in the case that a circularized accretion disc has formed at those times, the story cannot be as simple as previously described if such accretion flows are responsible for the optically bright events. 

There are three primary mechanisms that might convert the early bare disc emission to that with an SED peaking closer to the UV or optical. These are 1) Trapping of radiation in outflowing material by electron scattering, and the subsequent adiabatic losses of radiative energy as the photons do work on the outflowing material
2) absorption of disc photons by material at larger distances, which emits primarily at longer wavelengths 3) Compton scattering of photons off thermal electrons (here the electron thermal motions take the place of bulk kinetic motion that applied to point 1).

All three ought to be happening to some extent in the picture of a rapidly circularized accretion flow, and the manner in which they combine depends on the underlying hydrodynamics assumed. 
%Cartoon with disc + wind as Lodato and Rossi 2011
\begin{figure}[]
\centering
\includegraphics[scale=0.2]{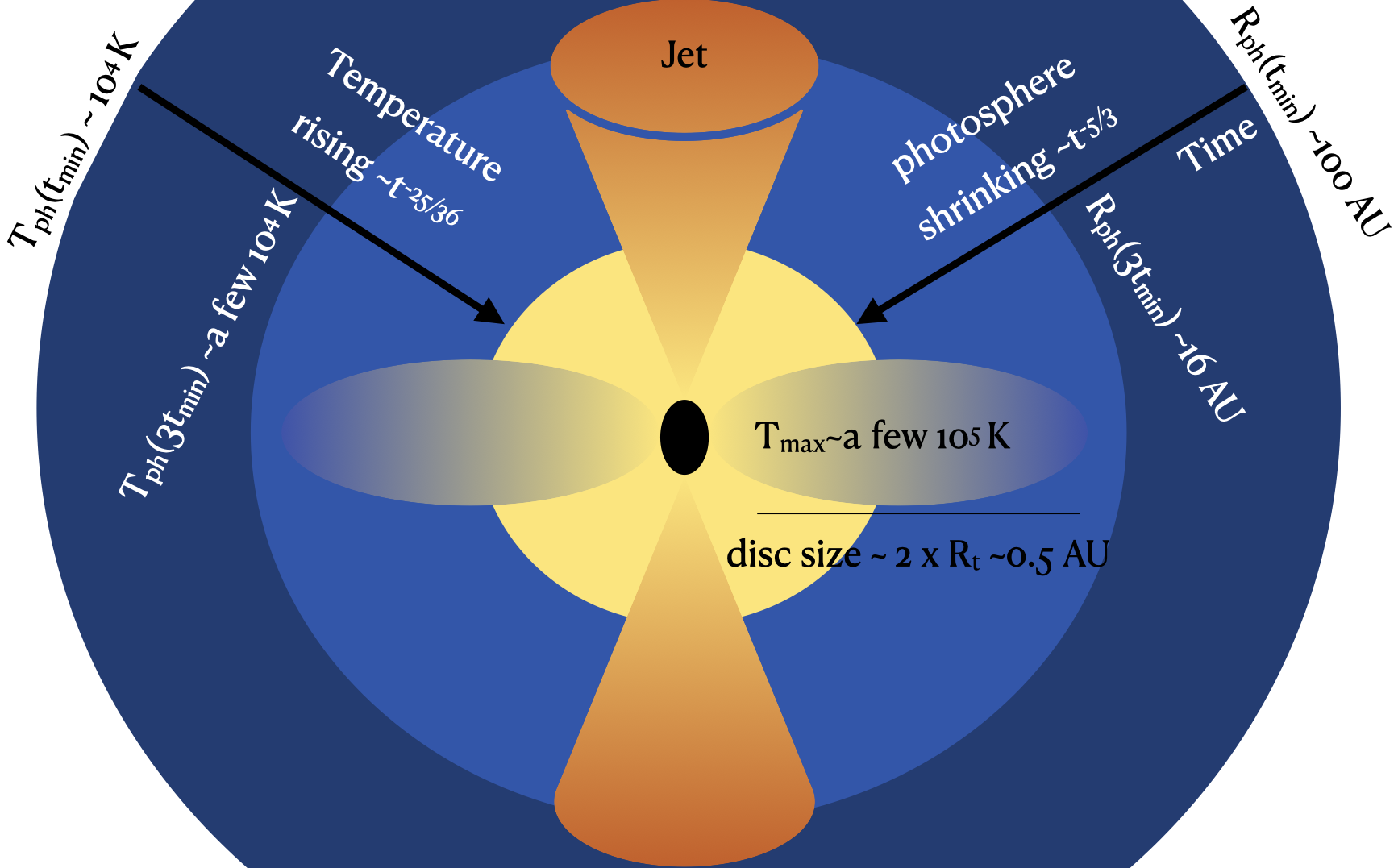}
\caption{Schematic of the disc-wind-jet structure that may arise following rapid circularisation of the stellar debris. The stated time dependencies of the shrinking photospheric radius and increasing photospheric temperature of the wind come from the adiabatic trapping model by \cite{Lodato:2011a}.}
\label{fig:disc_wind_cartoon}
\end{figure}

\subsubsection{Adiabatic trapping of radiation in expanding flows}
\label{sec:OutflowsAdiabatic}
As discussed earlier, when mass is brought into the vicinity of a black hole at a super-critical rate, an expected outcome is the formation of outflows that carry away mass, energy, and angular momentum that cannot be captured by the black hole. These outflows may be powered by the energy released during the initial collision of the bound debris with itself \citep[e.g.][]{Jiang:2016a,Sadowski:2016a,Lu2019}, or by energy released during the subsequent circularization and accretion of the circularized gas. In some cases the accretion-related outflows may take the form of a relativistic jet, either magnetically launched \citep[e.g.][]{Tchekhovskoy:2014a, Sadowski2015,Dai2018,Curd2019} or radiatively launched \citep[e.g.][]{Sadowski2015,Coughlin:2014a}. We do not focus on emission relating to the jet here, although we do discuss the radio afterglow signal associated with it in section \ref{sec:JetRadio}. Instead, here we focus on wild-angle, mildly relativistic or non-relativistic outflows.

The fundamental ingredients of the emission corresponding to these wide-angle outflows in the context of TDEs were laid out in SQ09, drawn in part from the treatment by \citet{Rossi:2009} for the disruptive tidal interaction of neutron star binaries by \citet{Rossi:2009}. The TDE outflow model has been subsequently refined by \citet{Lodato:2011a}, \citet{Metzger:2016b}, and \citet{Lu2019} (henceforth LR11, MS16, and LB19, respectively), and others. In short, the energy released from circularization and/or accretion goes into radiation that is trapped in a promptly launched, quasi-spherical envelope of outflowing material (see Fig.~\ref{fig:disc_wind_cartoon}). In the process of escaping, the radiation does work adiabatically on the gas, transferring energy from the radiation to the kinetic energy of the outflow. This causes the radiation that finally escapes to have a lower overall flux and peak emission frequency than it did at the point where the orbital energy of the gas was first converted to radiation. 

The various papers mentioned above differ on details such as the fraction of the falling back stellar debris that goes into the outflow, the location at which the outflow is launched, and whether the energy carried by the outflow comes only from the circularization process, or also from energy released from the accretion disc.  SQ09's and LR11's outflow models are driven primarily by circularization, and they predict the resulting bolometric light curve to rise as $t^{11/9}$, followed by a $t^{-5/9}$ decline that lasts until $t_{\rm cr}$ (equation~\ref{t_cr}).

 It is useful to keep track of a ``temperature'' $T_{\rm rad}$ corresponding to the radiation at all depths within the outflow. Here we use the term temperature not in a strict thermodynamic sense, but simply as a measure of the energy density of the radiation field: $T_{\rm rad} \equiv (E_{\rm rad} / a_{\rm rad})^4$, where $E_{\rm rad}$ is the radiation energy per unit volume, and $a_{\rm rad} \equiv 4 \sigma_{\rm SB} / c$. Of particular interest is the value of $T_{\rm rad}$ at the electron scattering photosphere, which we will label as $T_{\rm ph}$, since this will be related to the mean energy of the photons that escape the system. From the analysis of SQ09 and LR11, $T_{\rm ph}$ evolves gradually but does not stay constant, first declining as $t^{-7/36}$ and then, past the peak of the light curve, rising  as $t^{25/36}$. 

The predictions of MS16 are slightly different for several reasons, but most relevant here is that they keep track of when the radiation will no longer be sufficiently trapped in the outflow to continue to adiabatically transfer its energy to the gas. This estimate is made in terms of the so-called trapping radius $r_{\rm tr}$, the location in the outflow where the radiative diffusion time through the outlying material first equals the expansion time of the gas. This leads to the estimate that the optical depth beyond the trapping radius, $\tau(r_{{\rm tr}})$, equals $c/v_{\rm out}$, where $v_{\rm out}$ is the speed of the outflow, and where $\tau$ can be estimated by assuming that electron scattering is taken to be the dominant process that may trap the radiation. For a steady-state spherical outflow ejecting mass at a rate $\dot{M}_{\rm out}$, MS16 show that this leads to $r_{\rm tr} = \dot{M}_{\rm out} \kappa_{\rm es} / (4 \pi c)$, which evalutates to 
\begin{equation}
\label{eq:rtr}
r_{\rm tr} = 5.7 \times 10^{13} \,\, {\rm cm} \,\, \left(\frac{\dot{M}_{\rm out}}{1  \, \, M_{\odot} \,\, {\rm yr}^{-1} }\right) \left(\frac{\kappa_{\rm es}}{0.34 \, \,{\rm cm^2 \,\,g}^{-1}} \right) \, \, .
\end{equation}

This estimate assumes that the outflow corresponding to this steady $\dot{M}$ effectively extends to infinity - accounting for its truncation at finite radius, and the presence of material launched earlier at a different $\dot{M}_{\rm out}$, will modify the true location of  $r_{\rm tr}$. Indeed, while the mass outflow rate may remain approximately steady over several time intervals of $r_{\rm tr} / v_{\rm out}$, it will change over longer time scales. As the mass fallback rate drops, eventually $\dot{M}_{\rm out}$ in this model drops as well, and the trapping radius shrinks and eventually approaches the launching radius, at which time the radiation no longer loses a significant amount of energy radiative energy to adiabatic work. MS16 estimate this time, labeled $t_{\rm tr}$, by assuming that $\dot{M}_{\rm out}$ is a constant, near-unity fraction of the total mass fallback rate $\dot{M}$. They find $t_{\rm tr}$, to be only a factor of a few times $t_{\rm fb}$. Notice that the location of $r_{\rm tr}$ is important for determining how $L_{\rm bol}$ compares to the output from the bare disc.

By keeping track of how $r_{\rm tr}$ evolves over time, MS16 predict that after its initial rise, $L_{\rm bol}$ experiences a shallow decline similar to the  $t^{-5/9}$ predicted by the previous authors, while adiabatic losses are still important. Beginning around $t_{\rm tr}$, however, this then transitions to a decline tied to the standard approximation for the mass fallback rate, $t^{-5/3}$, as the accretion luminosity escapes without being adiabatically adjusted. Meanwhile, $T_{{\rm ph}}$ declines gradually at first, but after reaching its minimum when $L_{\rm bol}$ is at peak, $T_{{\rm ph}}$ rises as $\approx t^{1/2}$, which is more gradual than the $t^{25/36}$ rise in $T_{\rm ph}$ found by the previous authors. 

Once again, to connect to observations what is really needed is a prediction for the light curve in a particular band. Sometimes (e.g. LR11, MS16), the assumption is made that the radiation spectrum emitted from the scattering photosphere is fully thermalized and so can be represented as a blackbody emitting at $T_{\rm ph}$. This is not always justified. Even though the radiation is optically thick to electron scattering out to the photosphere by definition, that does not guarantee that it is optically thick to the absorption and emission processes that thermalize the radiation spectrum, as it would need to be at every wavelength of interest. 

We now list the predictions that can be made for these outflow models under the assumption that the radiation is emitted as a blackbody at $T_{\rm ph}$ after all. For LR11, for any band on the Rayleigh-Jeans tail  corresponding to $T_{\rm ph}$, then in this case $L_\nu$ will first rise as $t^{65/36}$ and then decline as $t^{-95/36}$. For MS16, for bands on the Rayleigh-Jeans tail, $L_{\nu}$ will initially rise and then remain nearly constant during the adiabatically trapped phase, but eventually transition to the rapid $t^{-95/36}$ shortly after $t_{\rm tr}$. In both cases, $T_{\rm ph}$ is typically high enough that optical bands lie in this Rayleigh-Jeans regime.

In sections \ref{sec:ReprocessingAllTypes} and \ref{sec:PartialThermFreefree} we will take a closer look at the requirements for re-thermalization of the radiation and the consequences of partial thermalization. However, we will now briefly consider how the radiation field might evolve in the absence of \emph{any} absorption, and is instead affected only through trapping in the fast-moving outflow by scattering. Likewise, we will momentarily take this scattering to be coherent in the rest frame of the fluid - the effects of thermal electron motion on the photon scattering process will be discussed in section \ref{sec:Comptonization} on thermal Comptonization.

In these pure-scattering conditions, adiabatic losses will reduce the mean energy of the photons as they scatter through the outflow. This results in a ``stretching'' of the spectrum so that it extends to lower energies, making the spectral shape wider than that of a blackbody with the same peak energy \citep{Roth2018,Dai2018}. The amount by which the mean photon energy shifts can be estimated by determining the fraction of radiative energy that is lost up to the trapping radius, although in reality photons continue to lose energy as they scatter even beyond that radius. Another way to do the estimate, valid in the regime when the radiative diffusion time is short compared to the timescales over which the outflow is evolving, is to use an integral of the velocity divergence \citep[equation 20 of][]{Roth2018}. 

In summary, the radiation escaping from such promptly launched, quasi-spherical outflows might be bright enough to match the peak optical luminosities of observed TDEs (i.e. peak $\nu L_{\nu}$ in the approximate range of $10^{42}$ to $10^{43}$ erg s$^{-1}$ in optical bands), although the optical luminosity predicted by some versions of these models is too faint by about an order of magnitude when compared to many observed events and for typical parameter choices (see Figure 1). The challenge of producing bright enough optical emission from these outflows is exacerbated when the radiation is not efficiently re-thermalized in the outflow. In any case, after several $t_{\rm fb}$ the optical light-curves from these outflows are expected to drop rapidly and become subdominant compared to more direct disc emission. 

\subsubsection{Reprocessing via absorption and re-emission, and the requirements for re-thermalization of radiation}
\label{sec:ReprocessingRequirements}
In the last section we saw that material launched in outflows can alter the radiation from the bare disc by adiabatically converting radiative energy to gas kinetic energy. Throughout that discussion, coherent electron scattering was the only interaction between the radiation and the stellar material. Now we consider the role of absorption and re-emission on re-thermalizing the radiation at a lower temperature.

The idea that a ``reprocessing envelope'', consisting of disrupted stellar material at large radii, might intercept X-rays produced from the inner accretion and re-radiate at lower temperatures to produce bright UV / optical emission was first put forth at least as early as \citet{Loeb1997}. The outflows discussed in the last section are one potential origin of such a reprocessing envelope. A thick accretion torus \citep[e.g.][]{Coughlin:2014a} might also serve this purpose. Yet another case, not discussed here, is reprocessing by the tail of bound and unbound debris. \citep{Bogdanovic:2004a}.

Dust that is present in the host galaxy or formed in the reprocessing envelope might also reprocess a large portion of the emitted radiation to infrared wavelengths \citep[first noted in][]{Kormendy1995}. We defer that topic to the \ehcochap\ and instead we focus on reprocessing envelopes generated in some manner by the stellar debris without \textit{in situ} dust formation, and we focus on the UV/optical emission.

In all cases, we are considering a large fraction of the material from the disrupted star concentrated within $\lesssim 1000 r_g$. This is irradiated by a hot inner source of temperature $\gtrsim 10^{5}$ Kelvin and peak luminosity in the range of $10^{44}$ to $10^{45}$ erg s$^{-1}$. In these conditions, we expect the continuum UV/optical spectrum that escapes to be formed in regions of appreciable scattering depth and high ionization. This influences the re-thermalization of the radiation. Let $\alpha^{es}$ and $\alpha_{\nu}^{abs}$ respectively denote the opacities from electron scattering and from all absorption processes for frequencies bewteen $\nu$ and $\nu + d\nu$. We define an opacity ratio $\epsilon_{\nu}$ as
\begin{equation}                                                        \label{eq:OpacityRatio}                                                 \epsilon_{\nu} = \frac{\alpha_\nu^{abs}}{\alpha^{es} + \alpha_{\nu}^{abs}}                                                      \end{equation}
An elementary result of radiative transfer theory is that the \emph{effective} optical depth to absorption is then given by $\tau_{\nu, \rm tot} \sqrt{\epsilon_{\nu}} $. Said another way, photons emitted at optical depths above the thermalization depth 
\begin{equation}
\tau_{\nu, \rm therm} \equiv 1/\sqrt{\epsilon_{\nu}}
\end{equation}
have an exponentially increasing chance of being absorbed rather than eventually escaping to the observer by scattering their way out. We define the thermalization radius $r_{\nu,\rm therm}$ as the radius corresponding to the thermalization depth, or equivalently the radius where the \emph{effective} optical depth to absorption for the given frequency interval is 1. An analytic expression for $r_{\nu,\rm therm}$ in the simplified case where the only absorption process is free-free absorption is presented in section \ref{sec:PartialThermFreefree}.

If and when the absorption opacities are large enough that $r_{\nu,\rm therm}$ coincides with the true photosphere (the average surface of last photon interaction for any process), and if this is true for $\emph{all}$ frequencies of interest, then we may apply the properties of blackbody emission. If the blackbody photosphere is spherically symmetric, then we have
 \begin{equation}
 L_{\rm bol} = 4 \pi r_{\rm ph}^2 \sigma_{\rm SB} T_{\rm eff}^4 \,\, ,
 \label{eq:BBemission}
\end{equation}
 where $L_{\rm bol}$ here refers to the luminosity escaping from the system (not necessarily equal to the $L_{\rm bol}$ from the bare disc), and $r_{\rm ph}$ is the photospheric radius. In this situation, $T_{\rm eff}$ will also correspond to  the observed color temperature. Near the peak of the light curve, and for typical color temperatures associated with optically bright TDEs (\optchap), the spherical blackbody model puts the photospheric radius at \begin{equation}
 \label{eq:rph}
 r_{\rm ph} = 4.2 \times 10^{14} \,{\rm cm} \,\, \left(\frac{L_{\rm bol}}{10^{44} \, \,{\rm erg \,\, s}^{-1}} \right)^{1/2} \left(\frac{T_{\rm eff}}{3 \times 10^{4} \, \,{\rm K}} \right)^{-2}. 
\end{equation}
If $r_{\nu, \rm therm}$ lies beneath the scattering photosphere but the effective photon path to absorption is sufficiently small compared to $r_{\nu, \rm therm}$ for all $\nu$ of interest, then we can treat the system using equation~\eqref{eq:BBemission} along with a color correction, as discussed in section~\ref{sec:EarlyDiskEvolution}. For even lower values of $\epsilon_{\nu}$, $r_{\nu, \rm therm}$ becomes significantly different for each $\nu$, or does not exist at all for some $\nu$. In this case the spectrum will no longer have a blackbody shape, as discussed in the last section and also in section~\ref{sec:PartialThermFreefree}.

In TDE models with rapidly circularized accretion discs, the assumption that there is a large effective optical depth to absorption at all UV/optical wavelengths poses theoretical challenges. On the one hand, the radiative transfer calculations of \emph{static} reprocessing envelopes performed by \citet{Roth:2016a} suggested that there can be sufficiently high effective absorption optical depth to provide the observed optical flux, although even then the radiation was not completely thermalized. This calculation also assumed that a large fraction of the stellar debris ($\gtrsim 0.1 M_{\odot}$) can be present all at once in a roughly spherical distribution contained within a radius of $\lesssim 10^{15}$ cm. What remains to be seen is whether the same required absorption optical depth is present in more realistic, time-dependent, hydrodynamically motivated situations that assume rapid circularization. If there is an accretion disc involved, it will be crucial to track how that disc builds up, spreads out and evolves in thickness over time, or in the case of an outflow, whether the required effective optical depth can be sustained for a long enough time before the debris becomes too optically thin. 

MS16 have provided estimates of the ``ionization breakout'' time, when the envelope becomes optically thin to the absorption of soft X-rays from the inner disc\footnote{The analytic estimates for soft X-ray absorption in MS16 and Roth 2016 were based on the He II photoionization opacity. However, further examination of the Roth et al. 2016 results by those authors have indicated that the dominant absorption process for the soft X-rays was the photoionization of oxygen, the only metal included in that calculation. This suggests that improved X-ray breakout estimates must account for opacities from oxygen, nitrogen, carbon, and possibly other species, rather than only helium.}, and find that for $M_{\rm BH} \lesssim 10^7 M_{\odot}$, the ionizing radiation may take several months to escape. Such an X-ray breakout may have been observed in ASASSN-15oi, although another interpretation is slow circularization leading to delayed disc formation \citep{Gezari:2017a}. In any case, if X-ray breakout is to happen at all, it must occur in such a way that the envelope remains effectively optically thick at UV/optical wavelengths at least until the time of breakout. This is because for UV/optical emission to be coming from re-thermalized radiation, we will need the effective optical depth to be large for \emph{both} the soft X-rays (to completely or partially absorb them), and for the UV/optical wavelengths (to produce a thermal or near-thermal spectrum at those wavelengths). Otherwise, we would also expect to see the optical flux drop off faster than usual around the time of X-ray breakout (in such a scenario the peak of the re-emitted emission would mostly likely be moving farther into the extreme ultraviolet). In ASASSN-15oi, a smooth power-law decline could connect the optical flux before and after the X-ray flare. This suggests either that the system did in fact remain effectively optically thick to optical wavelengths past the time the X-rays could escape, or that something other than the model described here is responsible for the optical emission.

In practice, UV/optical emission from TDEs is almost always fit to a spherical blackbody with temperature and radius specified by equation \eqref{eq:BBemission}. It is usually possible to produce a satisfactory match to the emission in the observed bands this way, despite the theoretical challenges we have described for equation \eqref{eq:BBemission} to apply. However, it is usually the case that the optical bands do not cover the peak of the spectrum for the assumed blackbody, so it is impossible to know whether the true spectrum deviates from a blackbody at extreme ultra-violet wavelengths.

A feature of such fits for UV/optical blackbody temperatures is that they either remain constant over several months, or to the extent that the temperature changes it seems to do so slowly (\optchap). Again, this statement is subject to some uncertainty given that we usually do not observe the peak of the spectral emission. This behavior of the fitted temperature can be contrasted with supernovae, which tend to exhibit temperatures that decrease with time as the ejecta expands in cools (\impostchap). The fact that TDEs do not seem to cool in the same way has become an important feature used to identify promising TDE candidates in optical surveys \citep[e.g.][]{Hung2018} The consequence of the slowly changing temperature and the rising then falling $L_{\rm bol}$ is that $r_{\rm ph}$ is inferred to move out and then in. Connecting this to the underlying hydrodynamics is an important task for models seeking to explain optical emission this way. Some authors use this $r_{\rm ph}$ evolution as evidence for an outflow \citep[e.g.][]{Strubbe:2015a}, and we will see similar behavior in the models discussed in section \ref{sec:Mosfit}.

\subsubsection{Comptonization}
\label{sec:Comptonization}

In many accreting black hole systems, inverse Comptonization of soft photons by hot electrons transfers energy from the electrons to the radiation field and produces the harder X-rays that are observed. Such processes are likely happening in TDEs with observed hard X-ray emission such as Swift J1644. For example, in \citet{Bloom:2011a}, an external Comptonization model was suggested to explain the hard x-ray emission from this event, in which softer photons produced primarily by the accretion disc and/or circularization shock are up-scattered by relativistic electrons in the jet. It was later suggested that the hard x-ray power-law in Swift J1644 could also be produced by mildly-relativistic, more thermalized electrons in the jet \citep{Lu2017}. Bright, beamed hard X-ray emission from inverse compton scattering is also produced in simulations with a relativistic jet and radiative transfer post-processing \citep[][hereafter C19]{Dai2018,Curd2019}. In fact, even in cases in which no highly relativistic outflow was produced, and only thermal electrons were considered, the C19 radiative post-processing suggests that a tail of hard X-ray emission will be produced and potentially detected by viewers at a wide range of inclinations, which is line with \citet{Lu2017}. 

It is also important to consider the role that direct (as opposed to inverse) Comptonization will play in shaping the spectrum of TDEs at a range of lower energies including soft X-ray and UV, which will often involve the transfer of energy from the photons to the electrons. While the bound electrons responsible for most of the absorption and re-emission of light beyond the accretion disc may be far out of LTE, it is generally reasonable to conclude that at the relevant densities the free electrons responsible for scattering the light are in LTE. This means they possess a velocity distribution described by a Maxwell-Boltzmann distribution and an associated temperature $T_e$, which in general might differ from other characteristic temperatures assigned to a region of gas, such as the radiation color temperature there.

As such, we can use the Compton $y$ heuristic for determining the importance of Comptonization. This quantity if formed by multiplying the average change in photon energy expected for each scatter by the expected number of times the photon will scatter before escaping the system:
\begin{equation}
y \equiv \frac{4 k T_e}{m_e c^2} \tau^2 \qquad {\rm when} \,\, \tau \gg 1
\end{equation}
 Values of $T_e$ outside a jet/funnel take on values that range from several $10^4$ to $10^5$ Kelvin, while the photons from the inner disc may be more energetic. For $T_e$ of $10^{5}$ Kelvin, $y$ evaluates to 0.7 for $\tau$ of 100. This implies a significant shift in the photon energy due to thermal motion of electrons during repeated scattering.
 
In addition to computing $y$, it is equally important to identify the thermalization depth for a particular wavelength, since the relevant quantity is really $y$ above that depth. In other words, photon absorption competes with Comptonization. For example, as seen in the radiative post-processing of \citet{Dai2018}, the majority of the soft X-rays produced in the disc close to the black hole were affected by Comptonization and shifted to lower energies. However, when viewed at disc inclinations close to edge-on, the optical portion of the spectrum remained relatively unaltered between test cases where Comptonization was included or not. This can be understood because the optical thermalization depths tended to be closer to the scattering photosphere than for the soft X-rays, as treated in this calculation. Therefore there was a relatively small value for $y$ for the optical photons emitted at larger radii; that is, those photons were able to escape with fewer scatterings.

\subsection{Simplified TDE models for light curve fitting}
\label{sec:Mosfit}
We conclude this section with one approach to connect TDE light curves to the mass fallback rate \citep[][hereafter M19]{Mockler2019}, which has been implemented using the MOSFiT astronomical transient fitting framework \citep{Guillochon2018}. Along with everything that has been written in this chapter up to this point, this model assumes that  circularization is rapid for the portion of the debris that forms an accretion disc. This model invokes the presence of a reprocessing layer (section \ref{sec:ReprocessingRequirements}), emitting as a spherical blackbody, to convert radiative energy emitted from the disc at primarily soft X-ray wavelengths into UV/optical emission, at least for several $t_{\rm fb}$ while the reprocessing layer remains optically thick.

\subsubsection{Motivation and method}
\begin{figure}[h!]
\centering
\includegraphics[scale=0.8]{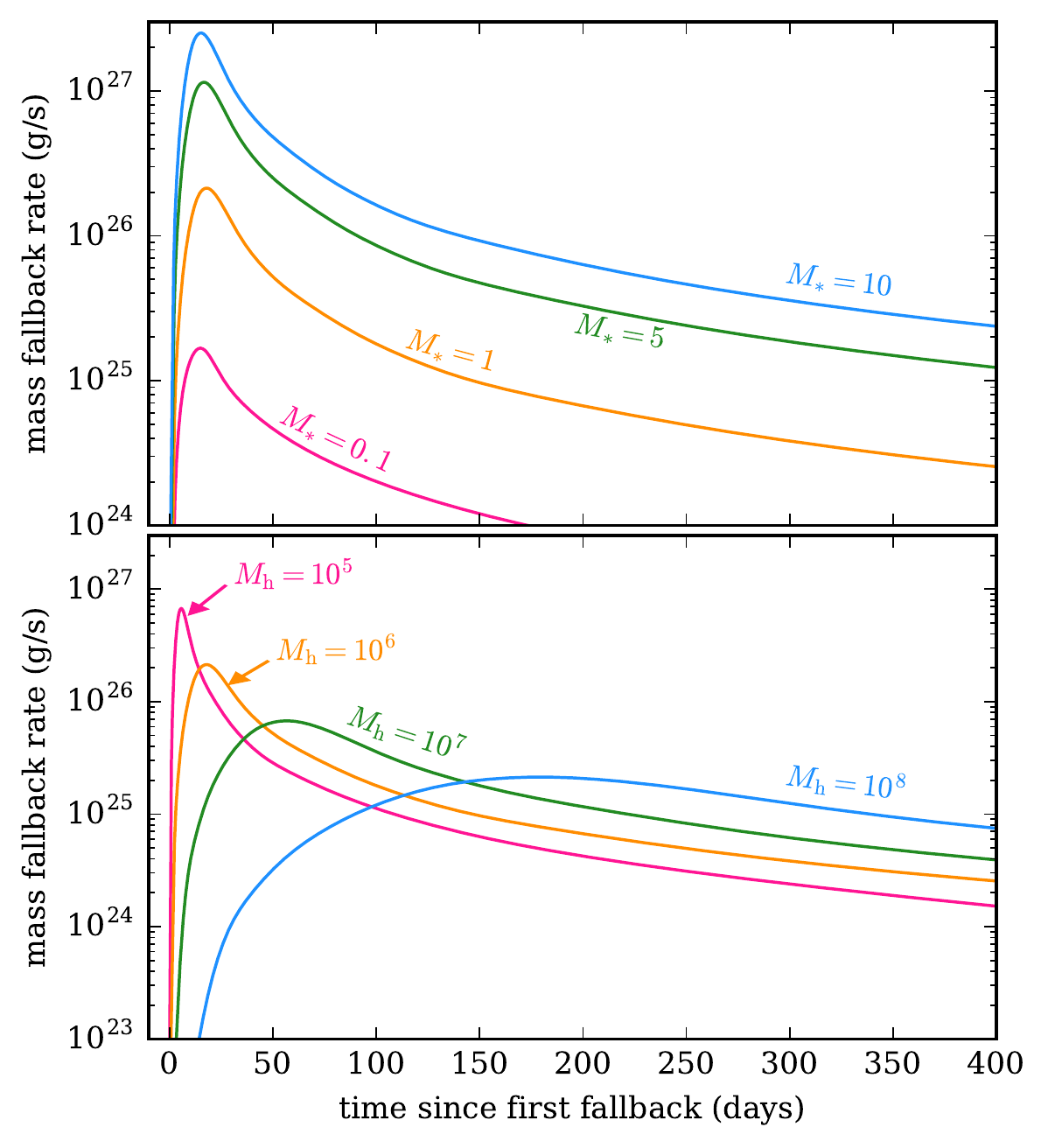}
\caption{Top panel: The mass of the disrupted star has a large effect on the overall normalization of the fallback curve, but a relatively minor effect on its shape. Bottom panel: The black hole mass, on the other hand, is largely responsible for setting the timescale $t_{\rm fb}$ of the fallback rate. If the observed light curves have a direct relation to the mass fallback rate, it may be possible to estimate $M_{\rm BH}$ based on the shape of the light curves.}
\label{fig:dmdt}
\end{figure}

Among the motivations for such an approach, we consider again simple estimates of the stellar mass fallback rate (see also section~\ref{sec:EarlyDiskEvolution}):
\begin{align}
{\rm peak} \, \, \dot{M_{\rm fb}} &\propto M_{\rm BH}^{-1/2} M_{\ast}^{2} R_{\ast}^{-3/2}, \nonumber \\
t_{\rm fb} &\propto M_{\rm BH}^{1/2} M_{\ast}^{-1} R_{\ast}^{3/2}. \nonumber
\end{align}
The black hole mass affects both the mass fallback rate and the timescale to the same extent. However, the mass-radius relationship of solar metallicity main-sequence stars is such that the effects of the star's mass and radius largely cancel in the equation for $t_{\rm fb}$. The timescale of mass fallback curves is therefore highly sensitive to the mass of the black hole, as shown in Figure~\ref{fig:dmdt}, even when simulations of the disruption process provide more accurate information than the simple scalings listed above. To the extent that the luminosity of observed TDEs is directly tied to the mass fallback rate as computed from hydrodynamics simulations \citep[e.g.][]{Guillochon:2013a}, fitting the light curves of TDEs may provide an indirect measurement of the mass of the disrupting black hole (M19). The mass of the star does have a second-order effect on the peak timescale and the power-law decline, and in some cases this is enough to fit for the stellar mass as well.  

\begin{figure}[htb]
\centering
\includegraphics[scale=0.8]{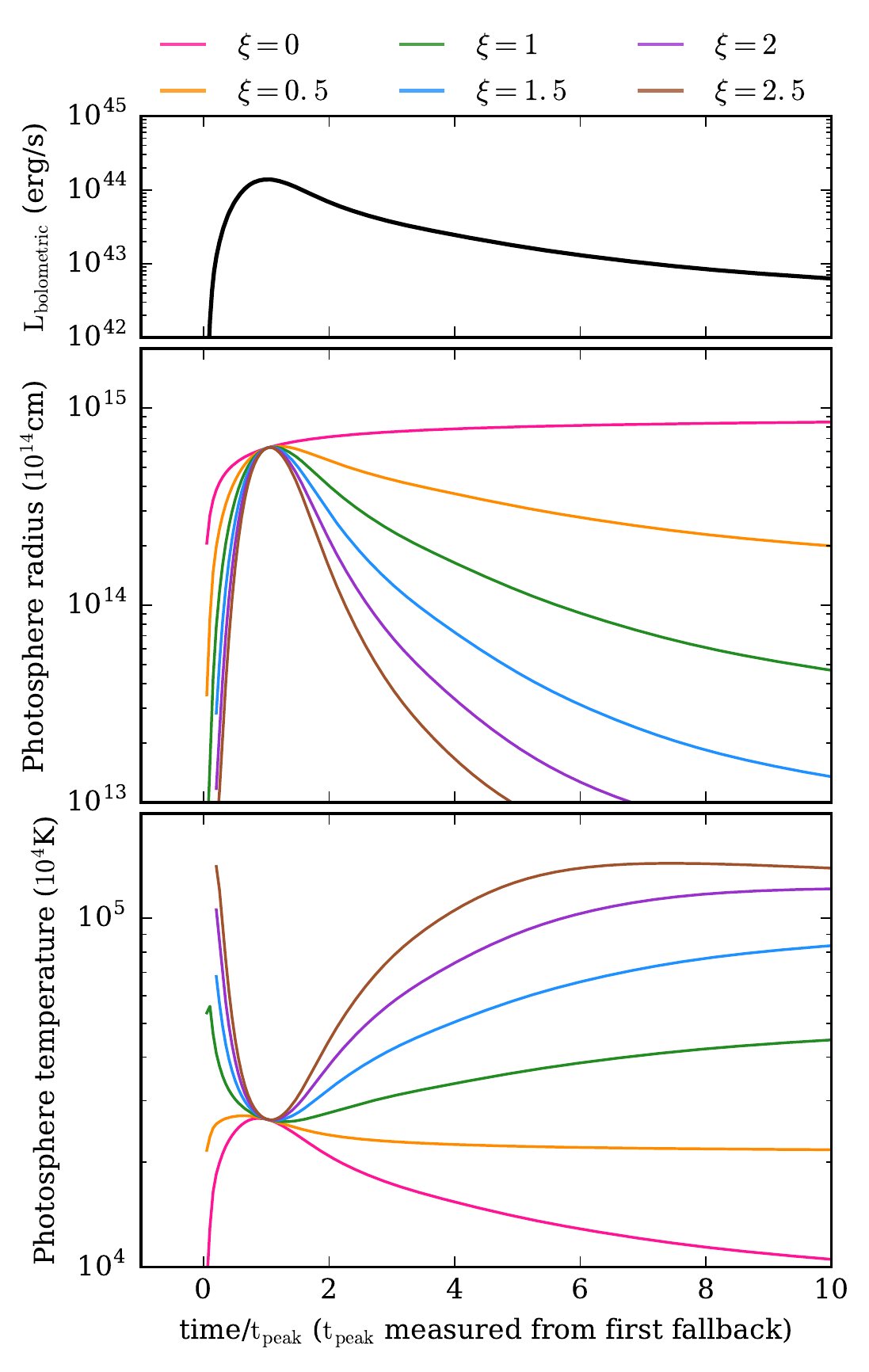}
\caption{Top panel: A reference bolometric light curve fit following the procedure of \citet{Mockler2019}, which assumes that it closely tracks mass fallback rates found in hydrodynamic simulations. Middle panel: the time evolution of the blackbody photospheric radius that is assumed to give rise to the UV/optical emission, for various choices of the tuning parameter $\xi$. Bottom panel: the evolution of the corresponding blackbody temperature. For $\xi \gtrsim 0.5$, the temperature first declines and then rises, as was the case for the outflow models discussed in section \ref{sec:OutflowsAdiabatic}}
\label{fig:LRT}
\end{figure}

As mentioned earlier in this chapter, the observed temperatures of TDEs do not change dramatically even while the luminosities change by many orders of magnitude, implying that such a blackbody photosphere is dynamic, growing and shrinking in radius in a manner that tracks the light curve. The precise physical origin of the reprocessing layer is left unspecified in what follows. However, given the dynamic nature of the photosphere, simulations of outflows launched during the circularization and/or accretion processes (section \ref{sec:OutflowsAdiabatic}) are especially relevant. In particular, \cite{Jiang:2016a} found that a solution of the form $R_{\rm phot} \propto L^{\xi}$ fit their simulations for a power law index of $\xi \sim1$, although it must be kept in mind that these simulations treated the stream intersection as an isolated system that did not interact with other stellar debris that might influence the evolution of the outflow over the course of the event. The same general power-law formula is used to model the photosphere in the MOSFiT model. M19 found that the best-fit power-law indices for a sample of 14 optical TDEs with bright optical and UV emission varied between $\xi \sim0.5$ and $\xi \sim 4$, with the majority of the fits preferring $\xi$-values between 1 and 2. For a blackbody photosphere to emit at a constant temperature as the luminosity increases, the power-law index $\xi$ would have to equal $0.5$. Both the simulations from \cite{Jiang:2016a} and the fitting results from M19 found $\xi > 0.5$, implying that the blackbody temperature decreases slightly near peak, and increases at late times as the effective photosphere radii decrease (Figure~\ref{fig:LRT}). This behavior qualitatively resembles the temperature evolution for the models discussed in section \ref{sec:OutflowsAdiabatic}.

\subsubsection{Sources of modeling uncertainty}

As discussed above, the effects of the stellar mass and radius on the TDE light curve largely cancel out when the disrupted star is young and has metallicity close to solar. This allows the peak timescale to be mostly sensitive to the mass of the black hole. However, varying the age and metallicity of the star changes the mass-radius relationship \citep{Choi2016}, allowing the mass and radius of the star to influence the peak timescale. This uncertainty can be expressed as a systematic error, as described in M19. They found that the error in the black hole mass due to this uncertainty is $\pm 0.2$ dex.

In this model it is often difficult to disentangle the influence of the mass of the star from that of the radiative efficiency. The main effect both of these parameters have on the light curve is to increase (or decrease) the bolometric luminosity, and therefore their degeneracy has very little impact on the uncertainty in the measurement of the mass of the black hole. 

Another source of uncertainty relates to the choice to set the radiative efficiency to a constant value for each event. Since the mass  fallback rate spans a large range of values, from possibly super-critical at early times to sub-critical at late times, we might expect the radiative efficiency to change significantly over time if the accretion rate is tied directly to mass fallback rate.

Most difficult to quantify are the other systematic uncertainties in the values of physical parameters derived such as the black hole mass using this method. These systematic uncertainties relate to the underlying assumptions that have been made to connect the fitting parameters to the underlying physical processes. Purely phenomenological models can be declared successes if they describe the data to within an acceptable level using some goodness-of-fit metric such as $\chi^2$. In fact, it is common to add pieces to such models until they meet this standard. The parameters may then be interpreted as actual physical measurements, but only when there is strong evidence for the underlying model.

As this book illustrates, there are many competing models for most aspects of these events. For this reason, a high priority of empirical studies will be to test and distinguish these models, in addition to exploring the values for physical parameters that are returned using an assumed model. Thus, it is important to keep in mind that even when a model may be able to reproduce the measurements at a statistical error level consistent within the error level of the data, there is more work that must be done to reduce the aforementioned systematic uncertainties.

\section{Emission from shocked tidal debris that does not rapidly circularize}
\label{sec:ShockedStreamsAll}
We now discuss a post-disruption hydrodynamic scenario that differs qualitatively from most of the models discussed in section~\ref{sec:RapidCirc}, which generally had assumed that an accretion disc containing a large fraction of the bound stellar material could form rapidly (compared to $t_{\rm fb}$) upon fallback. In section~\ref{sec:StreamHydro} we briefly review the underlying hydrodynamics that motivates a picture where the bound stellar material remains incompletely circularized for several months following the initial fallback. Sections~\ref{sec:StreamsInner} and ~\ref{sec:StreamsOuter} describe the inner and outer portions of this flow, which may give rise to prompt emission of soft X-rays and UV/optical emission, respectively. Section~\ref{sec:ShockedStreamsLightcurves} then applies these principles to infer the physical parameters of the event ASASSN-14li based on its multi-band emission.

\subsection{Underlying hydrodynamics}
\label{sec:StreamHydro}
It is common to pose the entire problem in terms of a characteristic scale, the nominal tidal radius $r_t \equiv R_* (M_{\rm BH}/M_*)^{1/3}$, defined by the distance from the black hole at which tidal gravity is comparable to the star's self-gravity.   However, this number is only an order-of-magnitude estimate of the radius at which strong tidal effects occur because its derivation is on the basis of dimensional analysis rather than actual dynamical calculation.

As first indicated by Newtonian simulations using stars with polytropic internal structures disrupted at a systematically ajusted range of impact parameters \citep{Guillochon:2013a}, and refined by more recent fully-relativistic simulations using stars with realistic main-sequence internal structures \citep{Ryu2019-1,Ryu2019-2,Ryu2019-4}, the critical radius within which a star is completely disrupted is in general a number of order unity times $r_t$ \citep[see also][for non-relativistic treatments of the disruption of realistic stellar models]{Golightly2019,Goicovic2019,Law-Smith2019}. Because, somewhat fortuitously, this critical distance is nearly independent of stellar mass for all masses $\lesssim 3 M_\odot$ \citep{Ryu2019-1,Ryu2019-2}, the dependence of this order-unity correction $\Psi$ on stellar mass and black hole mass can be factored into one piece that is a function only of $M_*$ and another that is a function only of $M_{\rm BH}$ \citep{Ryu2019-1}.

Similarly, the nominal tidal radius may be used to make an order-of-magnitude estimate of the maximum binding energy of the tidal debris, $\Delta \epsilon \equiv GM_{\rm BH} R_*/r_t^2$, but both structural and relativistic effects can case the actual maximum binding energy $\Delta E $ to differ from $\Delta \epsilon$ by an order-unity factor $\Xi \equiv \Delta E/\Delta\epsilon$.  Like $\Psi$, $\Xi$ can also be factored into one portion dependent on $M_*$ and another dependent on $M_{\rm BH}$.

When matter is initially torn off the star, its orbital eccentricity lies in the range $1 \pm 2\Xi (M_*/M_{\rm BH})^{1/3}$ and its semi-major axis is at least that of the most tightly-bound orbit,
\begin{eqnarray}                                                a_{\rm min} &=& [1/(2\Xi)] r_t (M_{\rm BH}/M_*)^{1/3} = [1/(2\Xi)] R_* (M_{\rm BH}/M_*)^{2/3} \\ \nonumber                          &\approx& 3.5 \times 10^{14} \Xi^{-1} (M_*/M_\odot)^{0.22} M_{\rm BH,6}^{2/3}\hbox{~cm} \\ \nonumber                             &\approx& 2400 \, \Xi^{-1} (M_*/M_\odot)^{0.22} M_{\rm BH,6}^{-1/3}\, r_g. 
\label{eq:amin}
\end{eqnarray}
Here we have adopted a single power-law fit to the main-sequence radius-mass relation $R_* \simeq 0.9 R_\odot (M_*/M_\odot)^{0.88}$ \citep{Ryu2019-2}.
 The orbital period of this most tightly-bound matter sets the characteristic timescale for
the evolution of the system, $t_{\rm fb}$ (section \ref{sec:EarlyDiskEvolution}), but note that the expression given in that section requires a correction factor $\Xi^{-3/2}$.

To join a nearly-circular accretion flow at radii $\sim r_t$ therefore requires, at the very least, a substantial loss of orbital energy. As simulations have shown, the ``nozzle'' shock, where the returning streams converge near the stellar pericenter, is usually a rather weak shock because the convergence speed is a fraction $\sim (M_*/M_{\rm BH})^{1/3} \sim 10^{-2}$ times the orbital speed \citep[the latter henceforth S15]{Guillochon:2014a,Shiokawa:2015a}. It can eventually create larger deflections, but only for a minority of the returning mass (S15). Additional shocks can be generated by relativistic apsidal precession, but for many parameter combinations the precession angle is small enough that these shocks occur near the orbital apocenters, where there is little kinetic energy available for dissipation \citep[S15, ][]{Dai:2015a}. By converting some orbital energy into heat, these shocks can increase $1-e$ by factors of a few, but that still leaves the orbits strongly eccentric \citep[for related results, see also][]{Hayasaki:2016a,Bonnerot:2016a}. In contrast, for stellar pericenters closer than $\approx 10r_g$, where the precession angle is large enough to bring the shock close to the pericenter scale \citep{Dai:2015a}, the orbital energy can be dissipated much more rapidly.  However, when the pericenter is that small, the probability of direct capture is larger than the probability of a tidal disruption, so such events are relatively rare \citep{Ryu2019-5}. Another possibility is that, even for somewhat larger initial pericenters ($r_p \gtrsim 15 r_g$), gas is launched from the initial stream intersection point and sent on converging trajectories toward the black hole, leading to secondary shocks and increased dissipation \citep{Bonnerot2020}. For a detailed review, see the \flowchap. For the remainder of this section we focus on the case where these secondary shocks are not present and the rate of orbital energy dissipation is too low to promptly circularize most of the bound stellar debris falling back until several intervals of $t_{\rm fb}$ have elapsed.
\subsection{Inner flow}
\label{sec:StreamsInner}

Although the shocks resulting from apsidal precession  in this case cannot deflect material onto orbits with a scale $\sim r_t$, there are other mechanisms to deposit mass in such a compact flow.  In particular, as just mentioned, S15 found that the nozzle shock can do so for $\sim 1/3$ of the bound gas.  However, this does not necessarily happen immediately upon the debris' first return from apocenter.  In the simulation of S15, it took $\approx 3t_{\rm fb}$ for the nozzle shock to build in strength before it deflected significant amounts of gas toward smaller radii. Matter drawn from the larger-scale accretion flow was then pushed inward at a roughly constant rate, $\sim 0.1\times$ the maximum mass-return rate, until $\approx 10t_{\rm fb}$, after which the inflow rate fell. 

The duration of these different periods and the fraction of the mass that quickly enters a small accretion disc depends upon $M_*$ and $M_{\rm BH}$. In particular, it should be noted that S15 used a mass ratio $M_*$ /$M_{\rm BH}$ of 780, which is small compared to more common values of $\sim 10^6$ for the tidal disruption of a main-sequence star of a super-massive black hole, and this mass ratio influences the hydrodynamics of the nozzle shock and the stream intersection shock. Nevertheless, it is reasonable to expect that whenever the apsidal precession angle per pericenter passage is sufficiently small (the precise value and its related parameter dependences still requiring further investigation), the timescale to divert mass to smaller radii are generically a few $t_{\rm fb}$ and that only a minority of the bound gas enters an accretion disc quickly.

At later times, inflow into this inner region depends increasingly upon the level of internal stresses in gas at larger radii, presumably due to the stirring of MHD turbulence by the magnetorotational instability \citep{Balbus1998}.

\subsection{Outer flow}
\label{sec:StreamsOuter}

\begin{figure}[]
\centering
\includegraphics[scale=0.25]{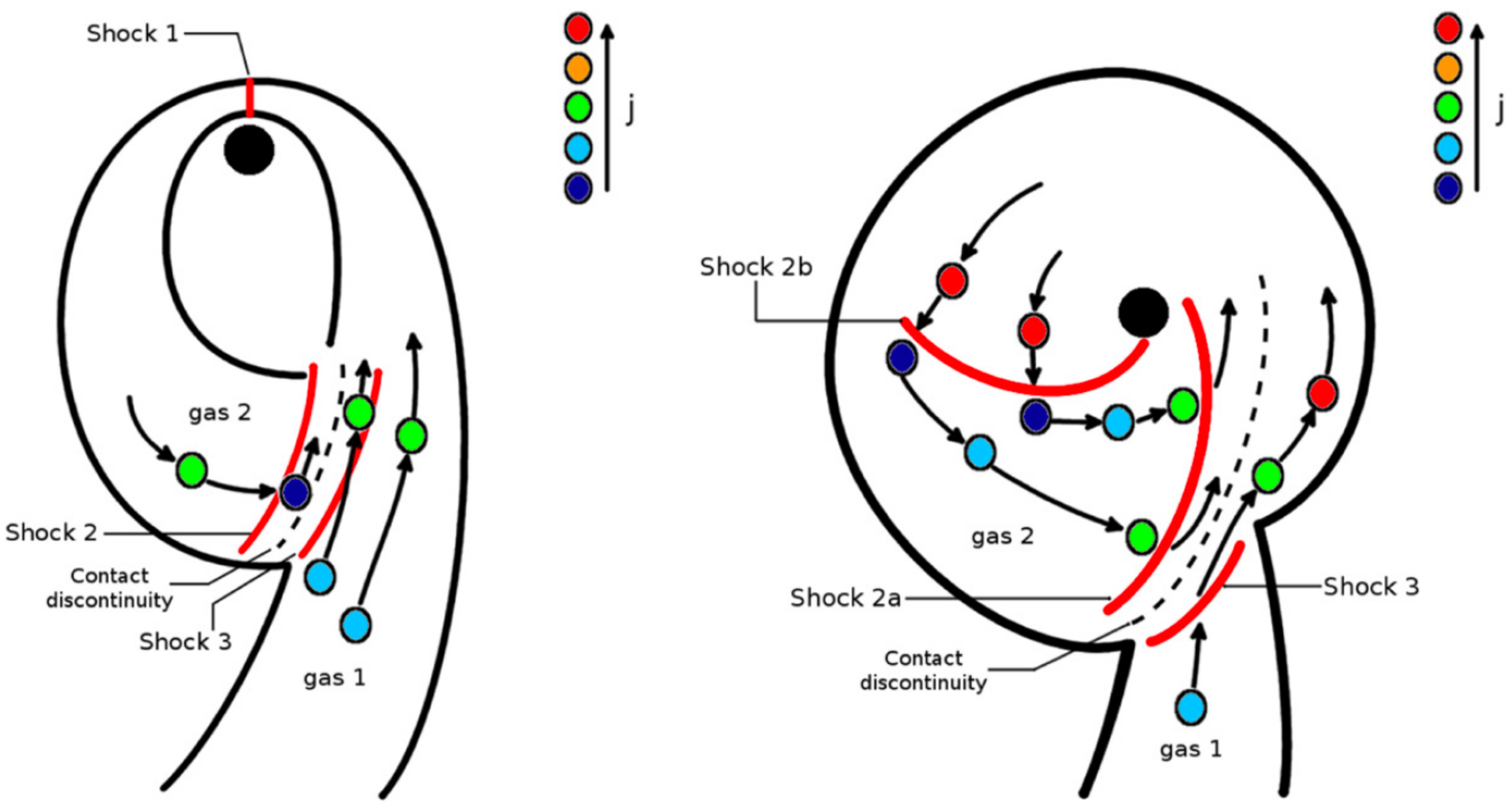}
\caption{A schematic for the bound stellar mass returning to pericenter and shocking with itself, as discussed in sections section~\ref{sec:StreamsInner} and ~\ref{sec:StreamsOuter}. This figure originally appeared in \citet{Shiokawa:2015a} and has been reproduced with permission. The ``outer flow'' consists of the majority of this bound stellar material that has returned and that remains on moderately elliptical orbits with semi-major axes on the order of $a_{\rm min}$ (equation~\ref{eq:amin}). The ``inner flow'' consists of the subset of the material depicted in the right panel that has been deflected by shocks into orbits of lower angular momentum and has dissipated a substantial fraction of its orbital energy. The color coding represents the specific angular momentum of the sample streamlines that have been drawn with arrows. For illustrative purposes, not all aspects of the diagram are drawn to scale.}
\label{fig:S15_fig12}
\end{figure}
By contrast, the fate of most of the bound mass after it has returned from its first apocenter passage is a very extended, highly eccentric, and rather asymmetric flow, as depicted in Figure~\ref{fig:S15_fig12}. ``Very extended''
means that its radial scale is comparable to that of the most-bound debris, $a_{\rm min} \sim 10^2 r_t$. Spreading half a stellar mass over an area $\sim \pi a_{\rm min}^2$ produces a Thomson optical depth (measured from the midplane vertically outward)
\begin{equation}\label{eq:optdepth} \tau_T \sim 500 \, \Xi^2 (M_*/M_\odot)^{0.56}  M_{\rm BH,6}^{-4/3}. \end{equation}
If the shocks, taken all together, have been able to magnify the streams' binding energy by a factor of a few, the internal energy density of the gas is comparable to the binding energy. In the outer portion of the orbit, where the matter spends most of its time, the vertical scale height would then be comparable to the local radius relative to the black hole. As a result,
the characteristic volume density is 
\begin{equation}                        \rho_* \sim 4 \times 10^{-12} \, \Xi^3 (M_*/M_\odot)^{0.35} M_{\rm BH,6}^{-2} \hbox{~gm~cm$^{-3}$},   \end{equation}
while the vertical component of gravity is
\begin{equation}
g_z \sim 1000 \, \Xi^2 (h/r) (M_*/M_\odot)^{-0.44} M_{\rm BH,6}^{-1/3}\hbox{~cm~s$^{-2}$}.                                  \end{equation}
Thus, the conditions at the surface of the bound stellar mass are much like those of stellar atmospheres.  The interior is sufficiently optically thick to be close to thermodynamic equilibrium, and the volume density near the photosphere is a few orders of magnitude smaller than in main sequence stars, while the photospheric surface gravity is roughly in the range of evolved massive stars \citep{Langer2014}.

Also like stars, a heat flux emerges through the surface of this flow.  However, the energy supply for this heat flux is not derived from nuclear reactions, but from shocks between the debris streams and, potentially, from the dissipation of MHD turbulence.  As shown by  \cite{PiranOptical2015}, the total heating rate due to shocks near the apocenter scales like $GM_{\rm BH}{\dot M}_0 /a_{\rm min}$, where ${\dot M}_0$ is the peak mass-return rate. Calibrated by the simulations of S15, the peak expected heating rate is 
\begin{equation}                         {\dot E} \simeq 1 \times 10^{43} \, \Xi^{5/2} (M_*/M_\odot)^{0.44} M_{\rm BH,6}^{-1/6} \hbox{~erg~s$^{-1}$}.
\end{equation}
In the simulation of S15, it lasted from $\approx 3t_{\rm fb}$ to $\approx 8t_{\rm fb}$ after the star's pericenter passage. Although $\dot E$ is $\sim 0.1 L_{\rm Edd}$ for fiducial values of the scaling parameters, it is
important to recognize that this is a small fraction of the luminosity expected from radiatively efficient accretion onto a black hole at the peak of the mass-return rate, only $\sim 10^{-3} \, \Xi (M_*/M_\odot)^{-0.22} (\eta/0.1)^{-1} M_{\rm BH,6}^{1/3}$.  Like the durations of peak shock heating, the calibration to the simulation of S15 may also be subject to some dependence on $M_*$ and $M_{\rm BH}$. If, for a different choice of system parameters, secondary shocks are formed at radii smaller than $a_{\rm min}$ such as those found by \citet{Bonnerot2020}, this would raise the heating rate above this estimate.

Spread over both sides of a flow occupying an area $\sim \pi a_{\rm min}^2$, this heating rate corresponds to a mean effective temperature
\begin{equation}                         T_{\rm eff} \simeq 2.2 \times 10^4 \, \Xi^{9/8} M_{\rm BH,6}^{-3/8}\hbox{~K}.               \end{equation}
This temperature is quite insensitive to parameters; by numerical coincidence, the main-sequence mass-radius relation leads to it being essentially {\it independent} of $M_*$, while it depends only modestly on $M_{\rm BH}$.  If the absorption opacity  in the gas is high enough that the effective mean free path to absorption at all UV and optical frequencies is small compared to $a_{\rm min}$, this effective temperature becomes a good estimator of the actual thermodynamic temperature of the surface. The case in which thermalization is not complete is treated in section~\ref{sec:PartialThermFreefree}.

Whether the heating emerges with a smooth surface brightness depends on whether the photon diffusion time from deep inside the flow is long or short compared to the time required for fluid elements to travel far from a shock front.  The latter time is generically $\sim t_{\rm fb}/2$; the former is
\begin{equation}
t_{\rm cool} \sim \tau_* (a_{\rm min}/c)(h/r) \simeq 6 \times 10^6 \, \Xi (h/r)  (M_*/M_\odot)^{0.78} M_{\rm BH,6}^{-2/3}\hbox{~s}.               \end{equation}
Thus, the cooling time and the gas orbital time are generally similar, with the orbital time perhaps a bit shorter; the ratio
\begin{equation}
t_{\rm cool}/t_{\rm fb} \sim 2 \Xi^{5/2} (M_*/M_\odot)^{-0.13} M_{\rm BH,6}^{-1/6}.
\end{equation}
As a result, the surface brightness (or, equivalently, the effective temperature) may or may not be thoroughly smoothed out by slow photon diffusion, but the degree to which it is smoothed is dependent primarily on $M_*$ through its determination of $\Xi$. Whatever its distribution, the profile of effective temperature due to shock heating does not resemble what might be expected from a classical accretion disc: there is no reason to expect it to be a declining power-law in radius, nor is it likely to be axisymmetric.

\subsection{Multi-wavelength properties and parameter inference: ASASSN-14li, a worked example}
\label{sec:ShockedStreamsLightcurves}

As previously argued, when relativistic apsidal precession is too weak to cause a sizable amount of the debris' orbital kinetic energy to be dissipated, the matter spreads out into a large ($r \sim (M_{\rm BH}/M_*)^{1/3} r_t$), eccentric, and asymmetric disc (S15). In this flow, the initial heating is predominantly due to shocks, both because the dynamics naturally lead to shocks and because MHD turbulence requires $\sim 10$~orbits to grow.  The optical/UV lightcurve during this initial phase is determined by the heating rate due to these shocks convolved with delays caused by photon diffusion through the matter.  For times beyond $\simeq 5$--$10t_{\rm fb}$ after disruption, the shock heating should decay $\propto t^{-5/3}$ because its ultimate energy source is the returning matter.  At later times, accretion from the apocenter region down to the black hole grows in importance to optical/UV emission on a timescale $\sim 10t_{\rm fb} \sim 1 \, \Xi^{-3/2} (M_*/M_\odot)^{0.32} M_{\rm BH,6}^{1/2}$~yr.

Although the simulation of S15 describes only a single, rather special, choice of parameters, its results in regard to the time evolution of the heating rate in these shocks can be checked for qualitative consistency with a large sample of observed events. For example, long-term HST monitoring of 12 TDEs demonstrated that in nearly all cases, the total energy radiated in the UV at times more than 1~yr after event-discovery was comparable to the energy radiated in that band within the first month \citep{vanVelzen2019}.  It immediately follows that the long-term lightcurve in this band is much better described by $t^{-1}$ than $t^{-5/3}$, which is closely consistent with the prediction of the previous paragraph, taken directly from a scaling of the S15 simulation.

To illustrate how to use the theory described in section \ref{sec:StreamsOuter} for parameter inference, it is easiest to work through an example. ASASSN-14li is especially suitable for this purpose \citep{Krolik:2016a} because it is the best-observed
TDE so far, with extensive X-ray and radio observations augmenting optical/UV data.  The availability of this multi-wavelength data provides both additional consistency checks and a broader view of what may be possible in a TDE.

We begin with the shape of the optical/UV spectrum.  In ASASSN-14li, it was well-fit by a black-body spectrum with a temperature $3.5 \times 10^4$~K for as long as 600~d after this event's discovery \citep{Cenko:2016a,Holoien:2016b,Brown2017}. For fiducial parameter values this is less than a factor of 2 from the prediction of the apocenter shock model \citet{PiranOptical2015} given in Sec.~\ref{sec:StreamsOuter} that $T_{\rm eff} \approx 2 \times 10^4 \Xi^{9/8} M_{\rm BH,6}^{-3/8}$~K.   Because the temperature depends on $M_*$ only through $\Xi$, the temperature is primarily a constraint on $M_{\rm BH}$.

A second constraint may be obtained from the total optical/UV luminosity, likewise estimated in Sec.~\ref{sec:StreamsOuter} on the basis of heating in apocenter shocks.  Integrating over the full black body spectrum, even though only part was directly observed, \citet{Holoien:2016b} found that the luminosity from ASASSN-14li at discovery was $\approx 6 \times 10^{43}$~erg~s$^{-1}$, again quite similar to the fiducial prediction, $\simeq 1 \times 10^{43} \Xi^{5/2} (M_*/M_\odot)^{0.44}
M_{\rm BH,6}^{-1/6}$~erg~s$^{-1}$.  As this expression demonstrates, if the observed luminosity is close to the peak luminosity, it is primarily a constraint on $M_*$, except for the $M_{\rm BH}$-dependence hidden in $\Xi$.

A different measure of this consistency is the effective area required to radiate the observed
optical/UV luminosity at the observed temperature.  Describing this area in terms of a double-sided disc of radius $r_{\rm eff}$, \citet{Brown2017} showed that $r_{\rm eff}$
peaked at $\simeq 1600 M_{\rm BH,6}^{-1}r_g$ and slowly declined to $\simeq 500 M_{\rm BH,6}^{-1}r_g$ over a period of $\sim 2$~yr.  These values compare well with a picture in which the matter orbits on a scale similar to the semi-major axis of the most-bound matter's orbit $\approx 2400 \, \Xi^{-1} (M_*/M_\odot)^{0.22} M_{\rm BH,6}^{-1/3} r_g$, especially when allowance is made for the facts that the shocks reduce its orbital energy by order unity and not all of the area of the eccentric accretion flow may contribute to the observed light.

A third arises from the emission line profiles.  In terms of full width at half maximum, the various UV lines seen in ASASSN-14li ranged from 1700~km~s$^{-1}$ to 7700~km~s$^{-1}$ \citep{Cenko:2016a}, while the optical lines began with full width at zero intensity $\simeq 10,000$~km~s$^{-1}$ but these narrowed to $\simeq 5000$~km~s$^{-1}$ at late times \citep{Holoien:2016b}. Given this range, one cannot define constraints too tightly; instead one can merely test whether the posited source region would possess a range of velocities consistent with what is seen at the factor of 2 level. In the apocenter region, and assuming that forces from radial pressure gradients do not dominate over centrifugal support, the characteristic orbital speed is
\begin{equation}\label{eq:vorb}
v_{\rm orb}(r) \eqsim 6000 \, \Xi^{1/2} (M_*/M_\odot)^{-0.11} M_{\rm BH,6}^{1/6} (2a_{\rm min}/r-1)^{1/2}\hbox{~km~s$^{-1}$},
\end{equation}
squarely in the observed range.  The last term in the estimate incorporates the variation in speed around an eccentric orbit of semi-major axis $a_{\rm min}$ if the instantaneous radius of a fluid element is $r$. Comparison to the observed profile width requires specifying both the range in orbital speeds corresponding to the range in orbital semimajor axes and radius contributing, and allowing for inclination.

Because the dependence of $v_{\rm orb}$ on $M_*$ and $M_{\rm BH}$ is so weak (and is largely through the factor $\Xi^{1/2}$), equation~\ref{eq:vorb} gives an extremely robust prediction of the scale of line-widths to be expected.  That it is so readily consistent with the magnitude of the observed line-widths, despite the almost complete lack of parameter-dependence, is also evidence on behalf of the underlying model. For the same reason, however, the line-width is not very useful for parameter inference because it is more sensitive to factors beyond the scope of the model (extent of the radiating region relative to the semimajor axis, orbital inclination) than to the parameters one would wish to constrain.

Nonetheless, detailed fitting of the optical emission line profiles to a model in which they are radiated by a
highly elliptical disc (a more specific application of the apocenter shock model) found that in the case of PTF09djl they were well described by a disc with semi-major axis $1700 r_g$ and eccentricity 0.97 \citep{Liu+2017,Cao2018}, which would be well in line with the fiducial estimate for $a_{\rm min}/r_g$. 

The next step is to examine the light curve. In the optical/UV, the flux was roughly constant for the first $\sim 25$~d and then began a decline that roughly followed an exponential form for the next
$\sim 150$~d \citep{Holoien:2016b}.  At still later times, the decline decelerated: from $\sim 250$~d until $\sim 550$~d post-discovery, the flux fell by only a factor $\simeq 3$.  An alternative description for the period comprising the exponential decline and the first $\sim 50$~d of the later epoch is $L_{\rm opt,UV} \propto (t + 29)^{-5/3}$ if $t$ is the time in days since discovery \citep{Brown2017}, but it fits the data better than the exponential only for the period 175--250~d after discovery.  In the UV, the late-time light curve was even shallower: from $\simeq 400$~d to at least $\simeq 1250$~d, there was no change in flux to within measurement error, and this flux was only a factor
$\sim 30$ below the flux at 60~d \citep{vanVelzen2019}.

In many TDE, only optical/UV data are available.  However, for ASASSN-14li, radio and X-ray data were also obtained.  Remarkably, equipartition analysis of the many epochs of radio data showed that the length scale of the synchrotron-emitting gas grew at a very nearly constant rate per unit time: $\sim 12,000$~km~s$^{-1}$ \citep{van-Velzen2016Sci,Krolik:2016a,Alexander:2016a}.  Moreover, this speed matches
the predicted speed of the fastest edge of the unbound ejecta (see section ~\ref{sec:UnboundDebris}) quite well. Having this constant speed allows one to project back to the date at which the outflow was launched: 70~days before discovery.  If one further identifies the outflow with the unbound ejecta, this launch date is also the date of the actual disruption.

It was also possible to detect the flare in X-rays, enabling a much improved estimate of the bolometric luminosity.
\citet{Brown2017} showed that for the first 450~d after discovery, the luminosity in soft X-rays stayed very nearly equal to the optical/UV if both are estimated by integrating over their respective blackbody fits (for the X-rays, the temperature was $\simeq 68$~eV for the first $\sim 200$~d, but fell to $\simeq 56$~eV for the next $\sim 200$~d).  Thus, for that period the bolometric luminosity was roughly double either the optical/UV or the X-ray luminosity separately. Interestingly, however, the peak value of $L_{\rm bol}$ corresponded to almost exactly the Eddington luminosity of a $10^6 M_\odot$ black hole, and it fell below that level within a few weeks after discovery.  If the bolometric luminosity were primarily due to accretion onto such a black hole at a rate given by the mass-return rate, this requires the accretion to be extremely radiatively inefficient (see section \ref{sec:EarlyDiskEvolution}).

However, if the optical/UV luminosity is generated in the apocenter region as a result of shock heating, one would expect its power to be roughly steady for a period of several $t_{\rm fb}$ beginning $\sim t_{\rm fb}$ after the actual disruption (although the mean flux may be more or less steady over this time period, the details of shock evolution may create fluctuations at the tens of percent level).  At later times, as the source of energy to the apocenter shocks, the infall of late-arriving debris, diminishes, so would $L_{\rm opt/UV}$. At still later times ($>\sim 10 t_{\rm fb}$), accretion inflow supersedes shocks as a source of heat in this region and the zone just interior to it.  Once this happens, $L_{\rm opt,UV}$ would decline as the mass reservoir in the apocenter region is depleted.  Because the inflow time at these distances is $\sim \alpha^{-1} (h/r)^{-2} t_{\rm fb}$, the decay in $L_{\rm opt,UV}$ would occur on timescales at least $\sim O(10) t_{\rm fb}$, consistent with the noticeable flattening of the optical/UV light curve after $\sim 250$~d.

Combining the optical/UV peak luminosity and $T_{\rm eff}$ constraints leads to
\begin{eqnarray}
\Xi^2 (M_*/M_\odot)^{0.44} &\simeq 4.6\\
\Xi^3 M_{\rm BH,6}^{-1} &\simeq 3.4
\end{eqnarray}
Accounting for how $\Xi$ implicitly depends on $M_\ast$ and $M_{\rm BH}$ \citep[equations 10 and 11 of][]{Ryu2019-1} leads to $M_* \simeq 2.8 M_\odot$, and $M_{\rm BH,6} \simeq 1.4$.  The corresponding $t_{fb}\simeq 28$~d.
If so, discovery took place when the system was $\simeq 2.5 t_{\rm fb}$ past disruption; the optical and X-ray plateau lasted another $\simeq 1 t_{\rm fb}$.
Because the black hole mass is derived entirely from TDE properties, it is very encouraging that the value derived from the apocenter shocks model ($1.4 \times 10^6 M_\odot$) is close to the value ($\simeq 1.7 \times 10^6 M_\odot$) estimated from its bulge velocity dispersion \citep{Wevers:2017a}.

\section{Additional considerations for the UV/optical/near-IR continuum and line emission}
\label{sec:LinesAndContinuum}
In this section we discuss general principles relating to the UV, optical, and near-IR emission from TDEs, regardless of whether this emission originates from a circularized accretion disc (section \ref{sec:ReprocessingAllTypes}) or shock-heated streams in non-circularized flows (section \ref{sec:StreamsOuter}). 

The state of the emitting gas in TDEs requires a separate set of assumptions than those commonly used in modeling either the broad-line region (BLR) of active galactic nuclei, or the simplest treatment of supernova spectra. Unlike the BLR, the gas is dense enough to be far from nebular conditions. By this we mean that the system is electron scattering dominated, and line photons, even from non-resonance lines, may be reabsorbed after many scatterings. For these reasons, a photoionization calculation which does not make use of a solution of the radiative transfer calculation will face difficulties in self-consistently exploring the conditions of interest. At the same time, the strong radiation field generated by the shocked and/or accreting gas prohibits the establishment of local thermodynamic equilibrium (LTE), an assumption that is often made to simplify the calculation of early-phase supernova spectra. Finally, the production of X-rays leads to Comptonization that is not often important in the context of supernovae (section \ref{sec:Comptonization}). The situation for radiative transfer in TDEs is most analogous to non-LTE treatments of AGN accretion disc emission \citep[e.g.][]{Hubeny2000,Hubeny2001}, stellar atmospheres \citep[e.g.][]{Mihalas1978-1,Hubeny1988}, and super-luminous supernovae driven by central engines \citep[e.g.][]{Dessart2012}.

In section~\ref{sec:PartialThermFreefree} we consider how these conditions affect the properties of the UV, optical, and IR continuum from TDEs. In section~\ref{sec:LineStrengths} we perform a similar analysis for the UV and optical emission and absorption lines, focusing on what sets the strengths of these features with respect to the continuum emission at neighboring frequencies.

\subsection{Constancy of the UV/optical color temperature: the potential role of partial thermalization of radiation}
\label{sec:PartialThermFreefree}

As mentioned in sections~\ref{sec:ReprocessingRequirements} and \ref{sec:StreamsOuter}, if the effective \emph{absorption} optical depth through the emitting stellar debris is low, the continuum emission might not completely thermalize, even when the gas has high total scattering depth. Here we examine this situation in more detail, and we begin with the case where the dominant absorption opacity is due to free-free processes, which are expected to dominate for TDEs at near-infrared (NIR) and longer wavelengths. Later we will make qualitative statements about how a partially thermalized spectrum at UV and optical wavelengths might help to explain the constancy of fitted blackbody temperatures over several months in TDEs, although at those wavelenghs other absorption opacities dominate over free-free.

The depth-dependent thermalization we will describe is related to continuum formation in Wolf-Rayet wind atmospheres \citep[e.g.][]{Wright1975}, but here we are in the scattering-dominated regime. A similar analysis to what follows was performed by \citet{Illarionov1972} and more recently by \citet{Shussman2016}, \citet{Lu2019}, and \citet{Margutti2019}. This particular derivation follows the latter reference most closely.

Neglecting gaunt factors, the free-free emissivity (erg cm$^{-2}$ s$^{-1}$ Hz$^{-1}$ Sr$^{-1}$) is
\begin{equation}                                    j_{\nu}^{ff}  \propto  T_e^{-1/2}Z^2 n_e n_i \exp \left(-h \nu/k T_e \right),               \end{equation}
where $T_e$ is the temperature of the free electrons. Near the surface of the emitting region, the electron temperature tends to level off \citep{Hubeny2000,Roth:2016a}, so that it can be absorbed into the constant. In the NIR, $h\nu / k T_e \ll 1$ and so the exponential factor is approximately constant. Thus, the free-free emissivity will vary with position roughly according to $j_{\nu} \propto n_e n_i \propto \rho^2$, with the assumption that the material is highly ionized. While the bound electrons are coupled to the radiation field and are likely to be out of the thermal equilibrium, the \emph{free} electrons and the ions should have a thermal velocity distribution given the densities of interest, so that $j_{\rm \nu}^{ff} = \alpha^{ff}_{\nu} B_{\nu}$. Expanding the Planck function to first order in $h \nu/ k T_e$, we find $\alpha_{\nu}^{ff} \propto \rho^2 \nu^{-2}$.

While the discussion up to this point has been independent of an assumed gas geometry, at this point we will assume spherical symmetry, and future work may extend this argument to other geometries. We assume that near the surface of the emitting material the density is dropping as a power-law, $r^{-n}$, for some $n > 1$. Due to the ionization from the engine, electron scattering dominates the opacity. The electron scattering optical depth (integrated from the outside in) is then wavelength-independent and given at radius $r$ by
\begin{equation}                                    \label{eq:TauEs}                                    \tau_{\rm es} = \left(\frac{1}{n-1}\right)\,\, \rho_0 \kappa_{\rm es} \,\, r_0^n \, \,  r^{1 - n},
\end{equation}
where $r_0$ is some reference radius within the region where this power-law expression for the density holds, and $\rho_0$ is the density at that location. This expression assumes the material in the power-law effectively extends to infinity.

We seek an expression for the thermalization radius $r_{\nu,\rm therm}$ (see section \ref{sec:ReprocessingRequirements}), which we recall is the radius where the \emph{effective} optical depth for frequency $\nu$ equals 1. Labeling the associated total scattering optical depth as $\tau_{\nu,\rm therm}$, and making use of our expression for $\alpha^{ff}_{\nu}$, we have $\tau_{\nu,\rm therm}  \approx 1/\sqrt{\epsilon_\nu} \propto \rho^{-1/2}\nu^{1}$.
Substituting this into equation~\eqref{eq:TauEs}, solving for the radius, and restoring the proportionality factors for a pure hydrogen gas, we have
\begin{align}
\label{eq:RthermNuScaling}
%r_{\nu,\rm therm} = \left[ \frac{1}{n - 1} \left( \sqrt{\frac{2 \pi}{3 k_B m_e} }\frac{4 e^6}{3 m_e k_B c} \frac{1}{m_p \sigma_T}\right)^{1/2}\, \rho_0^{3/2} \kappa_{\rm es} r_0^{3n/2} T_e^{-3/4} \, \nu^{-1} \right]^\frac{2}{3n - 2},
r_{\nu,\rm therm} = \left[ \frac{2 \left( 2 \pi \right)^{1/4} }{n - 1} \left(3 k_B m_e \right)^{-3/4} e^3 \left( m_p \sigma_T c\right)^{-1/2} \rho_0^{3/2} \kappa_{\rm es} r_0^{3n/2} T_e^{-3/4} \nu^{-1} \right]^\frac{2}{3n - 2},
\end{align}
where $m_e$ is the electron mass, $m_p$ is the proton mass, $k_B$ is the Boltzmann constant, $\sigma_T$ is the Thomson electron scattering cross section, and $e$ is the elementary charge in Gaussian-cgs units. When $n = 2$ this yields
\begin{eqnarray}
r_{\nu, \rm therm} &=& 2.2 \times 10^{14} \,\, {\rm cm}  \left(\frac{\rho_0}{10^{-12} \,\,{\rm cm}} \right)^{3/4} \left(\frac{\kappa_{\rm es}}{0.4 \,\,{\rm cm}^2 \,{\rm g}^{-1}} \right)^{1/2}  \,\, \left(\frac{r_0}{10^{14} \,\,{\rm cm}} \right)^{3/2}\nonumber \\
&\times& \,\, \left(\frac{T_e}{5 \times 10^4 \,\,{\rm Kelvin}} \right)^{-3/8} \left(\frac{\nu}{3 \times 10^{14} \,\,{\rm Hz}} \right)^{-1/2}.
\end{eqnarray}
Accounting for mixed gas composition is straightforward, but changes the result only slightly.

We can obtain an estimate of the observed spectrum by integrating the emissivity
\begin{equation}
\label{eq:ObservedSpectrum}                         L_{\rm \nu} \approx 4 \pi \int_{r_{\nu,\rm therm}}^{\infty} j_{\nu} \left(4 \pi r^2\right)  dr.
\end{equation}
Using our assumed density power-law, and our scaling for $j_{\nu}^{ff}$, we then have
\begin{equation}                                    L_{\rm \nu} \approx \frac{\left( 4 \pi \right)^2}{2n - 3} \,\, j_\nu\left(r_{\nu,\rm therm}\right) r_{\nu,\rm therm}^3 \propto r_{\nu, \rm therm}^{3 - 2n}.
\end{equation}
Using equation~\eqref{eq:RthermNuScaling}, we finally have $L_\nu \propto \nu^{(6 - 4n)/(2 - 3n)}$,
and for convenience we define $\beta \equiv (3 - 2n)/(3n - 2)$. The full result is
\begin{eqnarray}                             
L_\nu &=& \frac{4 \pi}{2n - 3} \,\left(6.8416 \times 10^{38} \,\, {\rm erg}\,\, {\rm cm}^3\,\,{\rm K}^{1/2}\right) \left(0.0177 \,\, {\rm cm^5}\,\,{\rm K}^{3/2} \,\,{\rm s}^{-2}\right)^\beta \nonumber \\
&\times& \left(\frac{1}{m_p^3 \sigma_T}\right)^{\beta}\, \rho_0^{2 + 3\beta} \,\, \kappa_{\rm es}^{2 \beta} \,\, r_0^{n(2 + 3 \beta)}\,\,T_e^{-(1+ 3\beta)/2} \,\,\nu^{-2 \beta}\,\, \exp\left( \frac{-h \nu}{k T_e}\right).
\end{eqnarray}
For $n = 2$, and on the Rayleigh-Jeans tail where $h \nu/ k T_e \ll 1$, we have
\begin{eqnarray}
\nu L_\nu &=& 1.5 \times 10^{41} \,\, {\rm erg s}^{-1} \,\,\left(\frac{\rho_0}{10^{14} \,\,{\rm cm}} \right)^{5/4} \left(\frac{\kappa_{\rm es}}{0.4 \,\,{\rm cm}^2 \,{\rm g}^{2}} \right)^{-1/2}  \,\, \left(\frac{r_0}{10^{14} \,\,{\rm cm}}\right)^{5/2} \nonumber \\ &\times& \,\, \left(\frac{T_e}{5 \times 10^4 \,\,{\rm Kelvin}} \right)^{-1/8} \left(\frac{\nu}{3 \times 10^{14} \,\,{\rm Hz}} \right)^{3/2}\,\, . 
\label{eq:FinalPartialThermSpec}
\end{eqnarray}
The $L_\nu \propto \nu^{1/2}$ and $T_e^{-1/8}$ dependencies for $n = 2$ were also found by LB19. For shallow density profiles with $n$ approaching 2/3, this result predicts $L_\nu \propto \nu^{\alpha}$ with $\alpha \to  - \infty$; $\alpha$ equals 0 at $n = 3/2$, and for $n \to \infty$, $\alpha \to 4/3$. In all these cases ($n > 2/3$), this result predicts a slope that is shallower than the Rayleigh-Jeans law $\alpha = 2.$

Equation~\eqref{eq:FinalPartialThermSpec} can only be directly applied to NIR wavelengths where free-free opacity dominates. Nevertheless, we now consider the possibility that the UV and optical continuum spectra of TDEs might be described by an expression similar to equation~\eqref{eq:FinalPartialThermSpec} in the sense that 1) $F_\nu$ is a power-law in $\nu$ or a sequence of power-laws over frequency intervals, where the power-laws exponents are not directly connected to the overall UV/optical flux, and 2) the overall flux normalization is sensitive to the gas density $\rho$ but not very sensitive to the gas temperature or bolometric luminosity.  This is not the standard interpretation for the UV and optical spectra of TDEs. Instead, they are generally fit to portions of thermal blackbodies (see section \ref{sec:ReprocessingRequirements}) because these yield better fits, in the sense of a lower reduced chi-squared statistic, than a single power-law as suggested by equation~\eqref{eq:FinalPartialThermSpec}. 

However, the peaks of thermal spectra at the fitted temperatures of roughly 2 to 4 $\times 10^4$ Kelvin lie at unobservable extreme ultraviolet wavelengths. So, it is not yet entirely clear that a thermal fit is appropriate, either, because the observations are not probing the portions of the spectra that would be most constraining of their thermal nature. Purely thermal UV/optical spectra also raise other theoretical puzzles. As discussed in section \ref{sec:ReprocessingRequirements}, for the highly irradiated gas in a reprocessing region of a TDE accretion disc, it can be hard to explain how a high enough optical depth to absorption can be maintained to re-thermalize the radiation. Perhaps more importantly, whether or not the stellar debris circularizes rapidly or not, it is difficult to explain why the blackbody temperatures for the thermal fits tend to remain nearly constant over several months in many events, even as the overall UV/optical flux drops by an order of magnitude or more. This would require the position of the photosphere to co-evolve with the UV/optical luminosity in such a way as to maintain the nearly constant color temperature. Such behavior is possible, but currently there is no hydrodynamic explanation for it.

On the other hand, a continuum spectrum resulting from partial thermalization, similar to the simple one derived here using only free-free opacity, might pose a solution to some of these theoretical challenges. It accounts for the low absorption opacities at optical wavelengths that might apply in TDE conditions. It also predicts a spectral slope that is independent of the overall UV/optical flux as a function of time, because in this case the shape of the continuum only depends on the overall luminosity only to the extent that a high ionization state is maintained. Meanwhile, the normalization of the spectrum (i.e. the total UV/optical flux) is most sensitive to gas density, which might be expected to drop in time, but the power-law slope should mostly persist unchanged, modulo some weak dependence on $T_e$ and the shifting of the exponential break at $h \nu / k T_e$. 

If the partial thermalization models are ever to succeed to the point where they can be used to fit spectra, they will need to be improved beyond their current state. To make quantitative time-dependent predictions, they will need to be tied to a hydrodynamic model. Additionally, they will need a  more complete treatment of the relevant absorption opacities at bluer optical and UV wavelengths where bound-free opacities from metals (especially oxygen, nitrogen, carbon and iron) are expected to dominate over free-free. Bound-free opacity is more difficult to treat because it depends sensitively on the precise ionization state of the gas, which is coupled directly to the radiation. The calculations described in \citet{Roth:2016a} included bound-free opacities from hydrogen, helium, and oxygen, and found that the  optical continuum generally took the form of a power-law with $\nu L_\nu \propto \nu^{2}$ ($\lambda L_\lambda \propto \lambda^{-2}$), that gradually steepened moving from optical to UV wavelengths. Additionally, as the mass in the reprocessing envelope varied, this would change the UV and optical flux but not the spectral shape. However, these calculations were not tied to a hydrodynamically motivated, time-dependent model for the reprocessing envelope, and did not include opacities from other metals which are likely to be important. Time will tell whether the partial thermalization explanation of the UV and optical continuum spectra is correct, or whether the purely thermal explanation can succeed after all.

\subsection{Line strengths with respect to continuum}
\label{sec:LineStrengths}

Similar principles of emissivity and thermalization depth set the strength of TDE emission lines with respect to the continuum. Consequently, inferences of the relative abundances of elements in the stellar debris based on emission line ratios, as is done in nebular spectroscopy, must be treated with caution. As one example, \citet{Roth:2016a} showed how the ratio of the He II / H$\alpha$ emission line ratio in optical TDE spectra can vary substantially as the density and spatial extent of the line-emitting gas are varied, while keeping the composition unchanged.

\begin{figure}
       \centering
       \includegraphics[width=\textwidth]{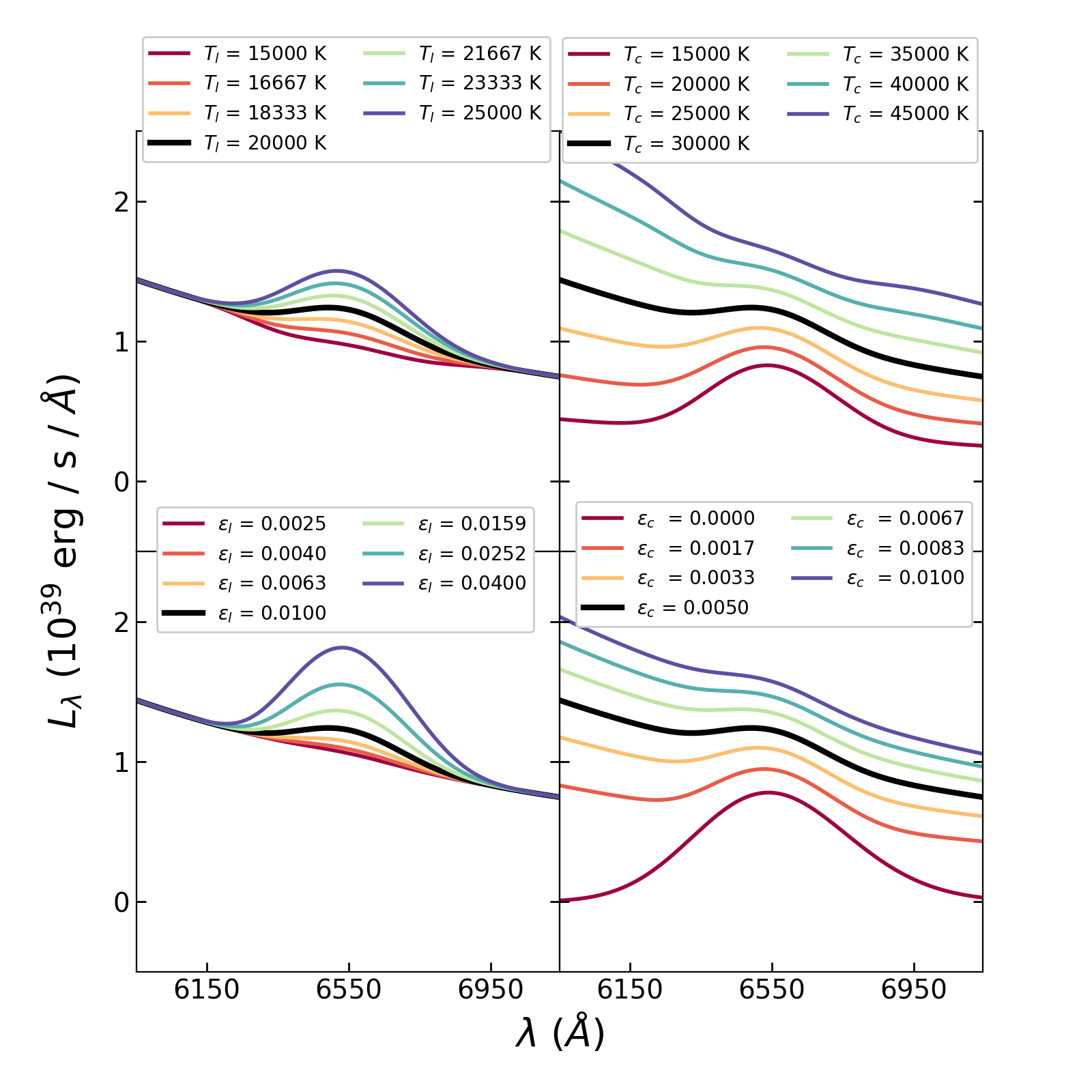}
       \caption{Line and continuum emission from the simplified plane-parallel atmosphere model described in this section. The emitting area is taken to be $(4 \pi) \left(10^{15} {\rm cm}\right)^2$. Depending on the opacities and emissivities of the line and nearby continuum, the equivalent width of the line may vary substantially. These effects must be considered when attempting to infer gas composition from emission line ratios in TDEs. The black emission line is the same in all four panels, and in each panel one parameter relating to the opacity or emissivity of the line or the continuum is varied.}
       \label{fig:line_strengths}
   \end{figure}

To gain more understanding of what sets these line strengths, we may perform the following toy calculation, which will culminate with Figure~\ref{fig:line_strengths}. We consider a plane-parallel atmosphere with surface area $A$ and we focus on a narrow range of wavelengths centered on a line. For concreteness we'll choose $H\alpha$ and its nearby continuum, although the formalism presented here also applies to other lines.

We adopt a Gaussian absorption profile $\phi(\lambda)$ for the line with Doppler width corresponding to a velocity of $10^4$ km /s . At the center of the line, we will denote the ratio of the line absorption opacity to the electron scattering opacity by $\epsilon_l$, and we will assume this line-center value is constant at all atmospheric depths we are considering. Similarly, the continuum has an opacity ratio denoted by $\epsilon_c$ that is taken to be constant with respect to both depth in the atmosphere and the range of wavelengths under consideration. 

We specify the source functions for these processes in terms of ``temperatures'' so that $S_{\lambda}(T) = B_{\lambda}(T)$ where $B_{\lambda}$ is the Planck function. ``Temperature'' is in quotes because in the radiation- and scattering-dominated environment of a TDE atmosphere the gas will not be in local thermodynamic equilibrium, so here temperature is merely serving as a convenient way to quantify the source function. The line and continuum each have their own source function temperatures, $T_l$ and $T_c$. These quasi-temperatures are distinct from the quasi-temperature $T_{\rm rad}$ introduced in section \ref{sec:OutflowsAdiabatic}. However, they come about for the same reason - the system is not in LTE. The closer the system gets to LTE, the closer all of these quasi-temperatures will come to equalling to the shared thermodynamic temperature.

In the limit $\epsilon_l, \epsilon_c \ll 1$, the combined source function for the line and continuum processes may be written as
\begin{equation}
  \label{eq:CombinedSourceFunction}
  S_{\lambda} = \frac{\epsilon_l \, \phi(\lambda) B_\lambda(T_l) + \epsilon_c B_\lambda(T_c)}{\epsilon_l + \epsilon_c}
\end{equation}
As discussed earlier, photons of frequency $\lambda$ that reach the observer will mostly be emitted from radii that satisfy $\tau_{es} < 1/\sqrt{\epsilon_{\lambda}}$ (here $\epsilon$ should be taken as the opacity ratio with contributions from both line and absorption processes). From equation~\eqref{eq:ObservedSpectrum}, with the understanding that the total emissivity requires the sum of the line and continuum emissivities, and that the thermalization depth depends on opacity contributions from both lines and continuum, this leads to the following estimate for the escaping luminosity \citep[c.f.][equation 1.103]{Rybicki1986}
\begin{equation}
  \label{eq:EstimatedLuminosity}
   L_{\lambda} \approx 4 \pi \sqrt{\epsilon_{l}\,\phi(\lambda) + \epsilon_c} \,S_\lambda A.
 \end{equation}
Figure~\ref{fig:line_strengths} displays the combined line and continuum emission as each of the parameters $\epsilon_l$,$\epsilon_c$, $T_l$, and $T_c$ are varied. Adjusting any of these affects the relative strength of the line with respect to the continuum. Additionally, these values will differ for each line in the spectrum and the continuum at wavelengths near that line. The values of these parameters can also be determined by a simultaneous solution of the radiative transfer equation and the non-LTE rate equations for the gas state. For rough comparison to observations, we have taken $A$ to equal $(4 \pi) \left(10^{15} {\rm cm}\right)^2$.   
 
 There are some general statements we can make. Ordinarily, the line-center opacity for H$\alpha$ (as measured by $\epsilon_l$) is greater than the continuum opacity, so that the line thermalizes closer to the surface than the continuum. Additionally, the continuum tends to rise at longer wavelengths in the optical due to the increasing free-free opacity at those wavelengths. However, to quantitatively determine the line ratios, the detailed calculation is needed. The range of values of $\epsilon_l$, $\epsilon_c$, $T_l$ and $T_c$ displayed in Figure~\ref{fig:line_strengths} share a broad overlap with the range of outcomes considered in \citet{Roth:2016a} for H$\alpha$; those calculations considered used reprocessing envelopes with total masses in the range $0.1$ to $0.5$ $M_\odot$, outer edges at radii between 5 $\times 10^{14}$ and 2 $\times 10^{15}$ cm, and bolometric luminosities between $5 \times 10^{43}$ and $10^{45}$ erg s$^{-1}$. However, it should be noted that the \citet{Roth:2016a} calculations tended to favor combinations with lower values of $\epsilon_c$ near H$\alpha$ (usually < 0.001) and with high enough values of $T_c$ to keep the line from contrasting too strongly with the continuum. Once again, it will be important to improve upon the assumptions and simplifications that went into those calculations before being able to use such a model to infer physical parameters from fitted spectra.
 
\section{Black hole environment: radio signals from unbound debris and jets}
\label{sec:BhEnvironsAll}

While the majority of the emitted radiation from TDEs arises from matter infalling onto the black hole, some radiation arises due to the interaction of unbound stellar debris with pre-existing gas in the circum-nuclear environment of the galaxy. There are two broad categories of this unbound material to consider: 1) the portion of the star that is unbound during its initial disruption, and 2) material that remains bound during disruption but is then launched and unbound during the subsequent circularization and/or accretion processes. This second class can be further classified into relativistic jets on the one hand, and the sub-relativistic outflows mentioned in section~\ref{sec:OutflowsAdiabatic} on the other. The material that is unbound during disruption will be present in every TDE where the star is fully disrupted, while the question of whether one or all of the other outflows are launched depends on the details of the subsequent hydrodynamics.

At least in two cases, the observed $\gamma$- and X-ray emission clearly indicate that a relativistic jet emerged \citep[Swift J1644:][]{Bloom:2011a,Burrows:2011a,Zauderer:2011a}, \citep[Swift J2058:][]{Cenko:2012b}. In these cases, aside from the prompt and beamed high-energy emission, the interaction of the jet with the surrounding environment is expected to be a source of radio emission \citep{Giannios2011}. Upper limits on this radio emission from samples of TDEs \citep{van-Velzen2013,Alexander:2017a} indicate that in many cases there is no evidence for environmental interaction with a jet. The expected properties of the radio emission produced in the forward shock of a relativistic jet when it \emph{is} launched, such as for the two cases highlighted above, will be discussed in section~\ref{sec:JetRadio}.

Aside from the events with jets, prompt radio emission has also been detected for a handful of other TDEs (\radiochap). Section~\ref{sec:UnboundDebris} will present an argument for how interaction of the highest velocity portion of the tail of unbound debris with the circum-nuclear environment might be responsible for these radio signals. Alternatively, these radio detections might be related to interaction of wide-angle, sub-relativistic outflows discussed in section~\ref{sec:OutflowsAdiabatic} with the circum-nuclear environment. For more details regarding that interpretation, please refer to the \radiochap.

\subsection{Radio emission from material unbound during disruption}
\label{sec:UnboundDebris}
The interaction of the unbound matter with the surrounding environment resembles  a supernova remnant, although the geometry is very different \citep{Guillochon:2016b,Krolik:2016a}. Depending on the adiabatic index, the unbound material may or may not fragment and collapse into dense blobs \citep{Coughlin:2016a}, or it many remain roughly wedge-shaped in the stellar orbital plane. In either case, it drives a bow shock in the ambient gas. Calculations such as those performed by \citet{Yalinewich2019-2} provide more details on the shape of the shock created by the unbound debris. Similar to a supernova remnant shock, this bow shock both accelerates electrons to relativistic velocities and amplifies the ambient magnetic field, producing a synchrotron-radiating region. This radio emission can vary significantly from one event to another as it depends sensitively on how much gas the debris plows through and the viewing angle. While self-gravity may play some role in confining the unbound debris and influencing its subsequent environmental interaction, this effect can be undercut when the ionized gas recombines, counteracting cooling \citep{Kochanek:1994a}. 

Originally, \citet{Guillochon:2016b} estimated that in typical galactic environments the unbound stream must travel a distance $> 1 $ pc, and possibly much farther, before sweeping up mass comparable to itself and appreciably decelerating. In this case the associated radio emission would be delayed by $\gtrsim 100$ years or longer. However, the hydrodynamic simulations of realistic stellar models performed by \citet{Ryu2019-1} found that for a large range of relevant values for $M_{\rm BH}$ and $M_\ast$, there is a larger spread in energy in the high-velocity edge of the unbound debris than previous estimates had indicated. This raises the possibility that a small amount of mass in the most energetic portion of the unbound debris stream can give rise to prompt radio synchrotron emission. 

The characteristic speed of the unbound ejecta at $R \gg a_{\rm min}$ is
\begin{equation}\label{eq:ejectaspeed} 
v_{\rm out} \simeq 6000 \hbox{~km~s}^{-1} M_{\rm BH,6}^{1/6} (M_*/M_\odot)^{-0.11} \Xi^{1/2},
\end{equation}
where once again we have used the approximate main-sequence mass-radius relation of $R_\ast = 0.9 R_\odot (M_\ast / M_\odot)^{0.88}$, as in section \ref{sec:ShockedStreamsAll}. This outflow carries a kinetic energy\footnote{See \citet{Krolik:2016a} for more refined expressions.} of
\begin{equation}
E_k \simeq 2 \times 10^{50} \,\, {\rm erg} \,\, M_{\rm BH,6}^{1/3} (M_*/M_{\odot})^{-1/3} \Xi \,\, .
\end{equation} 
As the ejecta continue to move out at nearly constant velocity they drive a bow shock in the ambient medium (at the relevant distances, deceleration from the gravity of the black hole is negligible).  The front edge of this outflow is determined by the small, but non-zero, fraction of the ejecta with greater energy: for stars with $M_* \gtrsim 1M_\odot$, $\sim 3 \times 10^{-4}$ of the unbound ejecta have speeds at least $2\times$ the fiducial value given in equation~\ref{eq:ejectaspeed} \citep{Ryu2019-2}.

To make a simple estimate of the synchrotron emission from the shock one can follow the same procedure as for calculating synchrotron emission from a supernova remnant \citep{Chevalier1998}, which is a Newtonian version of the calculation used for gamma-ray burst afterglows \citep{Sari1998}. At frequencies above the self-absorption frequency $\nu_A$, the spectrum will take on a power-law $F_\nu \propto \nu^{-(p-1)/2}$ where $p$ is the power-law index for the electron energy distribution, and often $p \sim 3.5$. Below this frequency, the spectrum will have slope $5/2$.

To present an estimate for $\nu_A$ and the emitted flux, we must define a number of quantities. The parameters $\epsilon_e$ and $\epsilon_B$ will represent the fraction of dissipated energy that goes into electrons and magnetic fields, respectively, while $\gamma_m$ represents the minimum Lorentz factor for the electron energy distribution. The geometric parameters $f_A$ and $f_V$ relate to the area and volume of the emitting region \citep[see][for details]{BarniolDuran2013-2}. The external medium into which the shock is expanding is taken to have a power-law density profile of the form $n(r) = n_0 (r / r_0)^k$. The near-constant velocity of the ejecta (prior to deceleration, which will occur when the shock sweeps up enough material to be comparable to its own mass) is represented by $v_0$. Then at time $t$ after the disruption, we find
\begin{eqnarray}                                        \nu_a &=& 0.4 \hbox{~GHz}\,\, f_A^{-2/7}  f_V ^{2/7} (\epsilon_e/0.1)^{2/7}  (\epsilon_B/0.1)^{5/14} (\gamma_m/2)^{2/7} \nonumber \\ &\times& (n_0/1500\hbox{~cm$^{-3}$})^{9/14}(r_0 /10^{16}\hbox{~cm})^{9  k/14} (t /100\hbox{~d})^{(4 - 9  k)/14} \nonumber \\  &\times& (v_{\rm o}/ 11,000 {\rm km/s})^{(14 - 9 k)/14} \ . \label{nua} \end{eqnarray}
and the corresponding flux measured at distance $d$ from the black hole, is 
\begin{eqnarray}                                                    F_\nu(\nu_a) &=& 3.4 \hbox{~$\mu$Jy} ~f_A^{2/7} f_V^{5/7} (\epsilon_e /0.1)^{5/7}  (\epsilon_B/0.1)^{9/14} (\gamma_m/2)^{5/7}  \nonumber \\ &\times& (n_0/1500\hbox{~cm$^{-3}$})^{19/14} (r_0 /10^{16}\hbox{~cm})^{19  k/14}  (t /100\hbox{~d})^{19(2-  k)/14} \nonumber \\ &\times& (v_{\rm o}/11,000\hbox{~km/s})^{(56- 19 k)/14} (d / 2.7  \times 10^{26}\hbox{~cm})^2 \ .  
\label{Fnua}                                                    \end{eqnarray}
\citep{Krolik:2016a}.   
 
These estimates made use of the faster velocities that are more likely to correspond to the fastest edge of the unbound debris, as discussed earlier. We also note that the flux may be higher for deeply penetrating (high $\beta$) events \citep{Yalinewich2019-2}.

\subsection{Radio emission from a relativistic jet and the radio emission from Swift J1644}
\label{sec:JetRadio}
The estimates of the radio emission due to the interaction of a relativistic jet with the surrounding matter  is similar to those derived for the radio afterglow from GRB jets \citep{Sari1998}. Like in the Newtonian case these estimates assume that a fraction of the blast wave energy is converted to magnetic energy and another constant fraction is converted to accelerated electrons. Once the kinetic energy of the jet and the external density are given the calculation of the synchrotron emission is straightforward. An essential generalization  is the consideration of a power-law density profile, as expected around a SMBH in a galactic center.  \cite{Giannios2011} find that for typical parameters the  synchrotron frequency is:
\begin{equation}                                        \nu_m =
\begin{cases}                       
  0.1 \hbox{~GHz}~  \left(\frac{\epsilon_e}{0.1} \right)^{2}  \left(\frac{\epsilon_B}{0.01}\right)^{1/2}   \left(\frac{n_0}{10\,\,{\rm cm}^{-3}}\right)^{1/2} & \hbox{forward shock,} \cr
  25 \hbox{~GHz}~ \left(\frac{\epsilon_e}{0.1}\right)^{2}  \left(\frac{\epsilon_B}{0.01} \right)^{1/2}     \left( \frac{n_0}{10 \,\,{\rm cm}^{-3}} \right)^{1/2} \left(\frac{\Gamma_j}{10}\right)^2 & \hbox{reverse shock ,}
\end{cases}
\label{num}                                             \end{equation}                                          where $\Gamma_j$ is the Lorentz factor of the jet. The corresponding maximal flux is:
\begin{equation}                                        F_{\nu_m} = 2 \hbox{~mJy}\,\, (\epsilon_j/0.01)  (\epsilon_B/0.01)^{1/2}   (n_0/10\hbox{~cm$^{-3}$})^{1/2} (\Gamma_j/10)^{-1} (D/{\rm Gpc})^{-2} \ ,                                \end{equation}
where $\epsilon_j$ is the ratio of the total kinetic energy of the jet to the total energy liberated by accretion.

The initial $\gamma$-rays and X-rays detected by Swift revealed the existence of a relativistic jet in Swift J1644 \citep{Bloom:2011a,Burrows:2011a}. As expected  \citep{Giannios2011} a few days after the event a radio signature was detected \citep{Zauderer:2011a}. However, the temporal evolution of the radio signal was not the expected one. While the initial expectation was that the energy involved would be roughly constant, it became clear that as time passed, more and more energy was added to the emitting region. This was clear from either a blast wave calculation of the kind mentioned above \citep{Berger:2012a,Zauderer2013,Eftekhari2018} or when an equipartition analysis was carried out \citep{BarniolDuran2013-1}. This latter analysis indicated that quite early on the source had slowed down and became mildly relativistic or even Newtonian. At the same time it showed that the energy within the emitting region increased by a factor of $\sim 10$ to 20 over $\sim$ 200 days. While the total amount of energy was within the total energy available in TDEs, it was not clear why the total energy increased as a function of time. One possible explanation was provided by \citet{Kumar2013}, who proposed that X-ray photons from close to the black hole are inverse-Compton scattered on the hot electrons within the radio emitting region, thereby cooling the electrons. As the X-ray flux decreased, this effect gradually disappeared, and more energy was available to the electrons to emit in the radio.

More information about radio emission from TDEs possessing relativistic jets can also be found in the \radiochap.

\begin{acknowledgements}
We thank Leiden Observatory BSc students Rens Verkade and Ashmara Wederfoort for providing Fig. 1. We thank the anonymous referees for providing feedback which greatly improved the quality of this chapter. NR acknowledges the support from the University of Maryland through the Joint Space Science Institute Prize Postdoctoral Fellowship
\end{acknowledgements}

% BibTeX 

% American Physical Society (APS) style, author-year citations
\bibliographystyle{aps-nameyear}      

% general.bib: this the file from James, that we use through the book, please don't add references to this file
% to add more references, that are not in general.bib, use extra.bib (and rename this file). 
\bibliography{general,extra}

\end{document}